\documentclass[prx,letterpaper,aps,10pt,superscriptaddress,twocolumn,floatfix,showpacs]{revtex4-1}
\usepackage{amsmath, microtype, graphicx, bm, amsfonts, hyperref, placeins, xcolor, dsfont}
\hypersetup{breaklinks=true}
\usepackage[noabbrev, capitalise, nameinlink]{cleveref}
\crefname{section}{Sect.}{Sects.}
\crefname{figure}{Fig.}{Figs.}
\crefname{equation}{Eq.}{Eqs.}
\crefname{appendix}{App.}{Apps.}

\newcommand*\diff{\mathop{}\!\mathrm{d}}
\newcommand{\hc}{\mathrm{h.c}}
\newcommand{\me}{\mathrm{e}}
\newcommand{\iu}{\mathrm{i}}
\newcommand{\mi}{\mathrm{i}}
\binoppenalty=\maxdimen
\relpenalty=\maxdimen

\usepackage{mathtools}
\newcommand{\commutator}[2]{\left[#1,#2\right]}
\newcommand{\anticommutator}[2]{\left\{#1,#2\right\}}
\DeclarePairedDelimiter\bra{\langle}{\rvert}
\DeclarePairedDelimiter\ket{\lvert}{\rangle}
\DeclarePairedDelimiterX\braket[2]{\langle}{\rangle}{#1 \delimsize\vert #2}

\DeclarePairedDelimiter{\abs}{\lvert}{\rvert}
\newcommand{\cross}{\times}
\newcommand{\matrixelement}[3]{\left\langle #1\middle\vert #2\middle\vert #3\right\rangle}
\newcommand{\expval}[1]{\left\langle #1\right\rangle}
\newcommand{\dv}[2]{\frac{\diff #1}{\diff #2}}

\begin{document}
\begin{abstract}
We theoretically investigate cooperative effects in cold atomic gases exhibiting both electric and magnetic dipole-dipole interactions, such as occurring for example in clouds of dysprosium atoms. We distinguish between the quantum degenerate case, where we take a many body physics approach and the quantum non-degenerate case, where we use the formalism of open system dynamics. For quantum non-degenerate gases, we illustrate the emergence of tailorable spin models in the high-excitation limit. In the low-excitation limit, we provide analytical and numerical results detailing the effect of magnetic interactions on the directionality of scattered light and characterize sub- and superradiant effects. For quantum degenerate gases, we study the interplay between sub- and superradiance effects and the fermionic or bosonic quantum statistics nature of the ensemble.
\end{abstract}

\title{Cooperative effects in dense cold atomic gases including magnetic dipole interactions}
\author{N. S. Ba\ss ler}
\affiliation{Department of Physics, Friedrich-Alexander Universit\"at Erlangen-N\"urnberg (FAU), Staudtstra{\ss}e 7,  D-91058 Erlangen, Germany}
\affiliation{Max  Planck  Institute  for  the  Science  of  Light,  Staudtstraße  2,  D-91058  Erlangen,  Germany}
\author{I. Varma}
\affiliation{Institut f\"ur Physik, Johannes Gutenberg-Universit\"at Mainz, 55122 Mainz, Germany}
\author{M. Proske}
\affiliation{Institut f\"ur Physik, Johannes Gutenberg-Universit\"at Mainz, 55122 Mainz, Germany}
\author{P. Windpassinger}
\affiliation{Institut f\"ur Physik, Johannes Gutenberg-Universit\"at Mainz, 55122 Mainz, Germany}
\author{K. P. Schmidt}
\affiliation{Department of Physics, Friedrich-Alexander Universit\"at Erlangen-N\"urnberg (FAU), Staudtstra{\ss}e 7,  D-91058 Erlangen, Germany}
\author{C. Genes}
\affiliation{Department of Physics, Friedrich-Alexander Universit\"at Erlangen-N\"urnberg (FAU), Staudtstra{\ss}e 7,  D-91058 Erlangen, Germany}
\affiliation{Max  Planck  Institute  for  the  Science  of  Light,  Staudtstraße  2,  D-91058  Erlangen,  Germany}
\maketitle

\section{Introduction}\label{sec:introduction}

Scattering of light off atomic gases necessarily involves aspects of quantum cooperativity, arising from the common, hybrid reaction of closely positioned and mutually coupled quantum emitters to the external stimulation. The optical response can be very complex, as it strongly depends on the gas density and temperature, the type of atoms comprising the gas, as well as on the strength of the driving field. For low density and high temperature, an independent scattering regime can be obtained where the gas response can be deduced from the single atom response \cite{Labeyrie99, Labeyrie03}. For higher density, weak excitation, and still high temperature, cooperative aspects such as super- and subradiance (spontaneous emission rates larger or smaller than that of an isolated particle) are emerging \cite{Dicke54, Kwong2015, Bromley2016, Inouye99, Inouye99, weiss19, jennewein18, deoliviera14, roof16, kaiser16, guerin16, Gross82,Scully04,Svidzinsky10,Keaveney12,Javanainen14,Kaiser09,Zhu2016,Javanainen2016,Pellegrino2014, Jenkins2016}. High driving powers lead then to non-linear optical effects, as the atomic transitions can saturate and atoms become fundamentally nonlinear elements. In all three described cases, a simple quantum optics approach suffices, based on the open quantum system formalism, where the atoms are treated as pseudo-spins $1/2$ (with transitions between ground and excited electronic orbitals) responding to an external stimulation.

\begin{figure}[b]
  \centering
  \includegraphics[width=0.9\columnwidth]{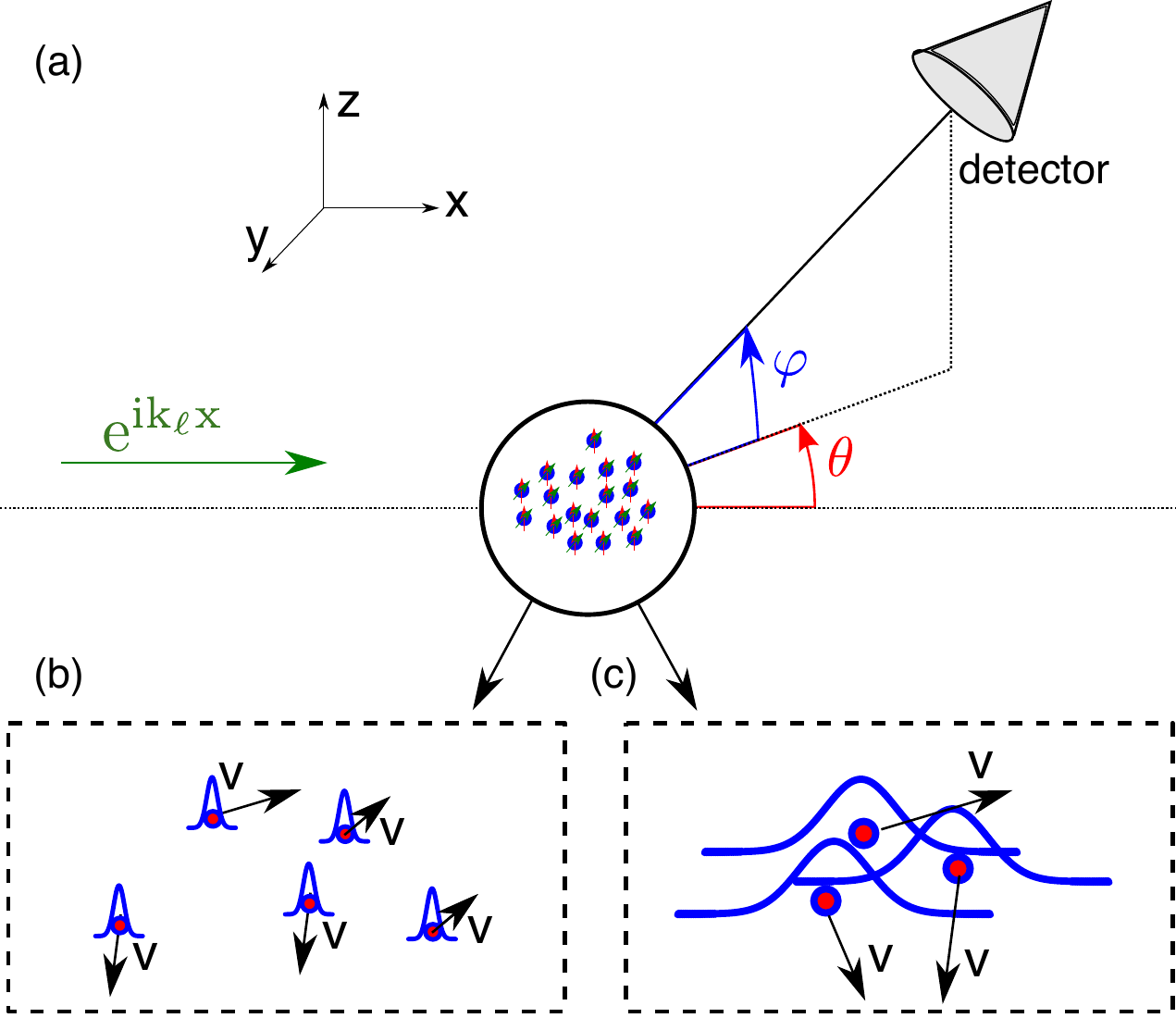}
  \caption {(a) An atomic gas is illuminated by a laser with frequency $k_\ell$ while a detector in the far-field at angles $\theta$ and $\phi$  measures the intensity of the scattered light. (b) In a first case, the de Broglie wavelength of each atom is much smaller than the average inter-particle separation, allowing one to treat the system as a classical thermal gas far from quantum degeneracy. The optical response can then be obtained in a simple fashion as the response of many coupled dipoles to external coherent stimulation. (c) In the opposite limit, quantum degeneracy is achieved by lowering the temperature and consequently producing particles with overlapping deBroglie wavepackets. To compute the optical response, a quantum many body approach to the problem is necessary.}
  \label{fig1}
\end{figure}

At low temperature, the motional wavepackets of the atoms comprising the gas can overlap, leading to a change in the theoretical framework, which must necessarily include a many body formulation to the problem \cite{Lewenstein94,Ruostekoski97,Javanainen99,Ruostekoski2016}. The two distinct limits, of classical versus quantum degenerate gases, are illustrated in Fig.~\ref{fig1}.

We analyze here a few distinct regimes and tailor our theoretical approach to each individual case, in order to deduce radiative emission properties. In a first step, for quantum non-degenerate gases, we extend the formulation of cooperative response to include magnetic interactions on top of the standard electric dipole-dipole exchanges, as for example strongly present in experiments with dysprosium atoms \cite{Childs1970, Petersen2020}. In the weak driving limit, the results show slight modifications in the superradiant response. In the strong drive limit, we show the emergence of a spin Hamiltonian with tunable parameters. In the next step, we assume an externally imposed potential and analyze the response of the quantum degenerate gas to external light drive in the case of both bosonic and fermionic statistics. Interesting aspects occur, as bosonic statistics implies superradiant behavior, similar to the standard Dicke superradiance example, even in the absence of any particle-particle interactions. In the fermionic statistics case, subradiance instead characterizes the emission properties of the gas.

The manuscript is organized as follows. In Sec.~\ref{sec2} we introduce the light-matter interaction model at the Hamiltonian level for both degenerate and non-degenerate gases including both, electric and magnetic, dipole-dipole interactions. In Sec.~\ref{sec3} we reduce our treatment to the non-degenerate case, where we first analyze the strong driving limit and show the emergence of a many particle spin Hamiltonian with interactions of tunable strength. We then exemplify the magnetic dipole interactions effect onto the light scattered in the weak driving regime and find small deviations from a purely electrically interacting gas. In Sec.~\ref{sec4} we analyze the light scattered from a quantum degenerate gas and find scaling laws for both bosonic and fermionic statistics. In Sec.~\ref{sec5} we describe an experimental platform based on thermal clouds of atomic dysprosium where the effects described above could possibly be tested. We conclude in Sec.~\ref{sec6}.

\section{Model}
\label{sec2}

We consider $\mathcal{N}$ atoms (positioned at $\bm R_i$ with index $i$ running from $1$ to $\mathcal{N}$) trapped in an external potential. Optical addressing of each atom (both by classical fields and the quantum electromagnetic vacuum modes) is achieved by coupling to its single valence electron, which we denote by its position $\bm r_i$ measured with respect to the position of the nucleus. The atomic cloud is illuminated by a laser with Rabi frequency $\Omega$, wavevector $k_\ell$ and frequency $\omega_\ell=c k_\ell$ propagating in the $x$ direction as depicted in Fig.~\ref{fig1}a). Two distinct regimes emerge: i) a \textit{classical} limit where atomic motion is described by a thermal distribution of velocities and ii) a \textit{quantum limit} where atomic motion is quantized and atoms become indistinguishable, thus the quantum degenerate case. The distinction between the two cases is illustrated in Fig.~\ref{fig1}b) and Fig.~\ref{fig1}c) where the overlap of individual atomic de Broglie wavepackets indicates the criterion for the transition between the two limits. In both cases we will be interested in spectroscopic quantities obtained from a detector positioned in the far field regime with angles $\theta$ and $\phi$ with respect to the incoming laser.

For simplicity of presentation we proceed with a set of approximations which are usually performed in treating light-matter interactions. We perform the dipole approximation which assumes that the size of the electronic orbital is negligible with respect to any relevant optical transition wavelength $\lambda$. In addition, we assume non-overlapping electronic orbitals between neighboring atoms. The electric and magnetic fields are quantized in a fictitious box of volume $\mathcal{V}$ and expressed in terms of photon creation and annihilation bosonic operators $\hat a_{\bm k,\bm \epsilon}$ with frequencies $\omega_k=c|\bm k|$ and commutators $\commutator{\hat a_{\bm k,\bm \epsilon_{\bm k}}}{\hat a_{\bm k',\bm \epsilon_{\bm k}'}^\dagger}=\delta_{\bm k,\bm k'}\delta_{\bm \epsilon_{\bm k},\bm \epsilon_{\bm k}'}$ (where $\epsilon_{\bm k}$ is the polarization for the mode $\bm k$). The free space quantization yields the electric and magnetic field operators
\begin{align}
\hat{\bm E}&=\iu\sum_{\bm k,\bm \epsilon}g_k\omega_k\epsilon_{\bm k}\left(\hat a_{\bm k,\bm \epsilon}\me^{\iu\bm k\bm R}-\hat a_{\bm k,\bm \epsilon}^\dagger\me^{-\iu\bm k\bm R}\right)\\
\hat{\bm B}&=\iu\sum_{\bm k,\bm \epsilon}\left(\bm k\cross\epsilon_{\bm k}\right)g_k\left(\hat a_{\bm k,\bm \epsilon}\me^{\iu\bm k\bm R}-\hat a_{\bm k,\bm \epsilon}^\dagger\me^{-\iu\bm k\bm R}\right),
\end{align}
with $g_k=1/\sqrt{2\omega_k \mathcal{V}\epsilon_0}$ being photon coupling strength. From here on the treatment is distinct for the two assumed limits of either a gas with a classical distribution of velocities or a quantum degenerate gas of indistinguishable atoms.
\subsection{Semiclassical gas approach}
We follow the standard quantum optics approach where the internal electronic dynamics of each atom is treated in terms of Pauli matrices. To this end, we restrict the dynamics of the electron to a ground state $\ket{g}_i$ and a single excited state $\ket{e}_i$, separated by frequency $\omega$. Notice that in practice, this assumption is a very good approximation for optically pumped atoms. An example based on dysprosium atoms is shown in Sec.~\ref{sec5} where the ground state is represented by a magnetic sublevel with $J=8$ and $m_J=-8$ and the excited state with $J'=9$ and $m_J'=-9$.

The Pauli matrices represent transitions in the electronic degrees of freedom and are defined as $\sigma_i=\ket{g}_i\bra{e}_i$. The coupling to electromagnetic waves occurs via either the transition electric $\hat{\bm d}_i^{\phantom{\dagger}}=\bm d_i\sigma_i^{\phantom{\dagger}}+\bm d^*_i\sigma_i^\dagger$ and magnetic $\hat{\bm \mu}_{i}^{\phantom{\dagger}}=\bm \mu_i\sigma_i^{\phantom{\dagger}}+\bm \mu^*_i\sigma_i^\dagger$ dipole operators or via the static magnetic dipole operator $\hat{\bm \mu}_{s,i}^{\phantom{\dagger}}=\bm\mu_{i,e}\sigma_i^\dagger\sigma+\bm\mu_{i,g}\sigma_i\sigma^\dagger$. The static components of the magnetic dipole are computed within the respective electronic state $\bm\mu_{i,e}=-g\mu_B\bra{e_i}\hat{\bm L}_i\ket{e_i}$ and  $\bm\mu_{i,g}=-g\mu_B\bra{g_i}\hat{\bm L}_i\ket{g_i}$ where $\bm L_i$ is the angular momentum operator for atom $i$, $\mu_B$ is the Bohr magneton and $g$ is the Land\'{e} factor. The transition dipole matrix elements are computed between orbitals $\bm d_{i}=-e\matrixelement{e_i}{\hat{{\bm r}_i}}{g_i}$ and $\bm\mu_{i}=-g\mu_B\bra{g_i}\hat{\bm L}\ket{e_i}$. Due to the fixed parity of hydrogen-like orbitals the static electric dipole moment must vanish which is not the case for the static magnetic dipole moment.
The resulting Hamiltonian
 \begin{equation}
 \label{Ham2}
      \mathcal{H}= \mathcal{H}_\text{at}+\mathcal{H}_\text{em}+\mathcal{H}_\text{el+mag}+\mathcal{H}_\text{drive},
 \end{equation}
is a sum over the free Hamiltonians of the atoms $\mathcal{H}_\text{at}$, the electromagnetic vacuum Hamiltonian $\mathcal{H}_\text{em}$, the electric and magnetic dipole coupling to the radiation field $\mathcal{H}_\text{el+mag}$, and the semiclassical drive $\mathcal{H}_\text{drive}$.\\
The first two terms are explicitly written as
 \begin{equation}
      \mathcal{H}_\text{at}+\mathcal{H}_\text{em}=  \omega\sum_i\sigma_i^\dagger\sigma_i^{\phantom\dagger}+\sum_{\bm k,\bm \epsilon}\omega_k\hat a_{\bm k,\bm \epsilon_{\bm k}}^\dagger\hat a_{\bm k,\bm \epsilon_{\bm k}}^{\phantom\dagger},
 \end{equation}
where we have set the zero of the energy at the ground electronic state level and ignored the zero point energy of the vacuum modes. The coupling of the atomic system to the electric and magnetic quantum fields is
 \begin{equation}
      \mathcal{H}_\text{el+mag}=  \sum_i\hat{\bm d}_i\cdot\hat{\bm E}(\bm R_i)+\left(\hat{\bm \mu}_{t,i}+\hat{\bm \mu}_{s,i}\right)\cdot\hat{\bm B}(\bm R_i).
 \end{equation}
The semiclassical drive is characterized by the Rabi frequency $\Omega$ and is written as
 \begin{equation}
      \mathcal{H}_\text{drive}=  \Omega\sum_i\left(\me^{-\iu {\bm k}_\ell \bm R_i}\sigma_i+\me^{\iu{\bm k}_\ell\bm R_i}\sigma_i^\dagger\right),
 \end{equation}
 where ${\bm k}_\ell=k_\ell\hat{e}_x$ and $\hat{e}_x$ is the unit vector in the x-direction. The Hamiltonian above is the starting point for the models analyzed in Sec.~\ref{sec3}.

\subsection{Quantum degenerate gas approach}
In the opposite limit of a quantum degenerate gas, it is more convenient to introduce a two-species model where field operators $\Psi^\dagger_{e,g}(\bm{R})$ create atoms at some position $\bm{R}$ with their electron either in the ground or in the excited state. The dipole moments can now be written in terms of field operators as the action of the Pauli matrices is now expressed by combinations of the field operators. For example, the operator $\hat \Sigma(\bm R)=\Psi^\dagger_{g}(\bm{R})\Psi^{\phantom\dagger}_{e}(\bm{R})$ is the field-theoretical equivalent of the matrix operator $\sigma_i$ where the particle location, which was previously denoted by an index, is now denoted by a continuous variable due to the replacement of localized scatterers by fields. The meaning is that an atom in the excited state is destroyed while another atom in the ground state is created in exactly the same position. The commutation relations are the standard ones $\commutator{\Psi^{\phantom\dagger}_{\alpha}(\bm{R})}{\Psi^\dagger_{\beta}(\bm{R})}_\zeta=\delta(\bm R-\bm R')\delta_{\alpha,\beta}$, where $\alpha,\beta\in\{e,g\}$ are species indices while $\zeta$ specifies commutation or anticommutation relations depending on the bosonic or fermionic nature of the gas. The single particle motional Hamiltonian is written as
\begin{equation}
\mathcal{H}_0=-\bm\nabla^2/2M+V_\text{ext}(\bm R),
\end{equation}
where $V_\text{ext}(\bm R)$ is an externally applied potential (assumed quadratic in the following) that allows for the use of a simpler notation using the trap basis states. Notice that, with a state-independent choice for $V_\text{ext}(\bm R)$, the motional wavefunctions are the same for the excited type and ground state type atoms.

In second quantization, the total system Hamiltonian is obtained as an integration of the Hamiltonian density. As a next step, the field operators can be expanded in a conveniently chosen basis. In particular, we will consider the trap basis defined by the eigenvectors $\mathcal{H}_0\phi_{\bm n}=\omega_{\bm n}\phi_{\bm n}$. Particle creation and annihilation operators can then be defined as
\begin{equation}
  \label{eq:trap_basis_def}
  \begin{split}
    \hat g_{\bm n}&=\int\diff{\bm R}\left[\phi_{\bm n}(\bm R)\right]^*\Psi_g(\bm R)\\
    \hat e_{\bm n}&=\int\diff{\bm R}\left[\phi_{\bm n}(\bm R)\right]^*\Psi_e(\bm R).
  \end{split}
\end{equation}
The operators can be interpreted in the following way: when $\hat g_{\bm n}^{\phantom\dagger}$ is applied to the vacuum, it creates an atom in the trap state $\bm n$ and in the electronic ground state.  Similarly, $\hat e_{\bm n}^{\phantom\dagger}$ creates an atom in the trap state $\bm n$ and in the electronic excited state. The fermionic or bosonic nature of the atoms is consistently taken into account by the commutation relations of these operators. The transition from free particles to trapped ones is performed by the tuning of the single particle trapping potential which in turn affects the shape of the trap basis eigenvectors. With this, the free Hamiltonian of the atoms can be written as
 \begin{equation}
      \mathcal{H}_\text{at}=  \sum_{\bm n}\omega_{\bm n}\hat g^{\dagger}_{\bm n}\hat g^{\phantom\dagger}_{\bm n}+\sum_{\bm n}(\omega_0+\omega_{\bm n})\hat e^{\dagger}_{\bm n}\hat e^{\phantom\dagger}_{\bm n}.
 \end{equation}
The electromagnetic modes Hamiltonian $\mathcal{H}_\text{em}$ is the same as before. The coupling of the atomic system to the electric and magnetic quantum fields is described by the following Hamiltonian
\begin{equation}
  \label{eq:trap_hamiltonian}
  \begin{split}
   \mathcal{H}_\text{el+mag} &=  \sum_{\bm n,\bm m}\left(\hat \Sigma_{\bm n,\bm m}^{\phantom\dagger}+\hat \Sigma_{\bm n,\bm m}^\dagger\right)\matrixelement{\bm n}{\bm d \cdot\bm E(\bm R)}{\bm m}\\
   & +\sum_{\bm n,\bm m}\left(\hat \Sigma_{\bm n,\bm m}^{\phantom\dagger}+\hat \Sigma_{\bm n,\bm m}^\dagger\right)\matrixelement{\bm n}{\bm \mu \cdot\bm B(\bm R)}{\bm m}\\
    &+\sum_{\bm n,\bm m}\left(\hat g^\dagger_{\bm n}\hat g^{\phantom\dagger}_{\bm m}\bm\mu_g+ \hat e^\dagger_{\bm n}\hat e^{\phantom\dagger}_{\bm m}\bm\mu_e\right)\cdot\matrixelement{\bm n}{\bm B(\bm R)}{\bm m},
  \end{split}
\end{equation}
where $\hat \Sigma_{\bm n,\bm m}^{\phantom\dagger}=\hat g^\dagger_{\bm n}\hat e^{\phantom\dagger}_{\bm m}$ are ladder operators destroying an excited atom in trap state $\bm m$ and creating a ground state atom in trap state $\bm n$.
The semiclassical drive is now written as
\begin{equation}
      \mathcal{H}_\text{drive}=  \Omega\sum_{\bm n,\bm m}{\hat \Sigma}_{\bm n,\bm m}^{\phantom\dagger}\eta_{\bm n,\bm m}+\hat \Sigma_{\bm n,\bm m}^\dagger \eta_{\bm n,\bm m},
 \end{equation}
where we have defined the Franck-Condon factors $\eta_{\bm n,\bm m}=\matrixelement{\bm n}{\me^{-\iu\bm k_\ell\bm R}}{\bm m}=\int\diff{\bm R}\left[\phi_{\bm n}(\bm R)\right]^* {\me^{-\iu\bm k_\ell\bm R}}\phi_{\bm m}(\bm R)$. For tight trapping conditions, where the localization of the atoms is on a level much smaller than the wavelength $2\pi/k_\ell$, the exponential can be approximated with unity and the Franck-Condon factors are equal to $\delta_{\bm n \bm m}$. In addition, to take into account the first-order correction, a Lamb-Dicke limit approximation can be made and only the matrix elements of the linear term $\iu \bm k_\ell \bm R$ would need to be considered.

\section{The classical motion limit}
\label{sec3}
Let us now focus on the semiclassical case, which assumes classical dynamics for the atomic motion, while the electronic dynamics is described in a quantum fashion in terms of Pauli matrices. As a function of the drive intensity, the emergent physics can be quite distinct. In a first case, where a high-intensity drive is assumed, high-excitation levels can be reached, with many atoms being excited at the same time. This is the standard regime for Dicke superradiance, i.e. the quick burst of spontaneous emission from an initial fully excited ensemble of closely spaced atoms. The other limit we consider is the weak-excitation limit, where there are hardly any excitations present in the system, rendering it possible to perform a transformation to a fully classical coupled dipole model. Despite its simplicity, the weak-excitation limit gives insights in the emergence of cooperative effects and their role in modifying directional scattering of light. In particular, we emphasize the role of magnetic dipole-dipole interactions and their competition with the widely studied electric counterpart.
\begin{figure}[t]
  \includegraphics[width=\columnwidth]{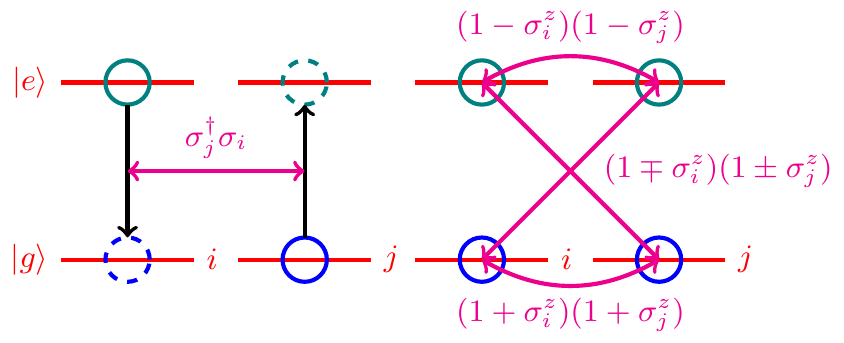}
  \caption{Illustration of interactions present in the effective Hamiltonian Eq.~\eqref{eq:effective_hamiltonian}. On the left is the excitation hopping due to the electronic dipole interaction between sites i and j. Initial occupations are indicated in filled open circles and the final occupations in dotted open circles. As indicated by the filled circle, the initially excited state on site $i$ is changed to the ground state by $\sigma_i$. Similarly, the ground state on site j, indicated by a filled circle is excited due to $\sigma_j^\dagger$. On the right side is the density-density interaction due to the static magnetic dipole interaction with the same color coding. Here, either the ground or excited state can be occupied on $i$ and $j$, leading to different interactions indicated by arrows between them and annotated with the appropriate interaction terms.}
  \label{fig2}
\end{figure}
%
\subsection{Effective spin Hamiltonian}
\label{sec:effective_ham}

From the Hamiltonian listed in Eq.~\eqref{Ham2}, one can derive a master equation for the evolution of the $\mathcal{N}$ electronic systems under the approximation of a frozen gas. Such approximation can hold for low enough temperatures, where motion evolution is slow compared to the time taken by any radiative processes. The derivation is based on the elimination of the photonic degrees of freedom~\cite{Lewenstein94, James93, Lehmberg70, Ruostekoski97, Javanainen99}, as outlined in detail in Apps.~\ref{sec:derivation}, \ref{sec:appendixb}, \ref{sec:appendixc}, \ref{sec:elimination}. The effective Hamiltonian describing dynamics in the reduced subspace of dimension $2^\mathcal{N}$ of the electronic degrees of freedom reads
\begin{equation}\label{eq:effective_hamiltonian}
  \begin{split}
    \mathcal{H}_{\text{eff}}&=\Delta\sum_i\sigma_i^\dagger\sigma_i+\sum_{i\neq j}\left(g^{(d)}_{ji}+g^{(\mu)}_{ji}\right)\sigma_i^\dagger\sigma_j^{\phantom{\dagger}}\\
    &+\Omega\sum_i\left(\me^{-\iu\bm R_i{\bm k}_\ell}\sigma_i+\me^{\iu\bm R_i{\bm k}_\ell}\sigma_i^\dagger\right)\\
    &+\frac{1}{8}\sum_{i\neq j}\Omega^{e,e}_{i,j}(1+\sigma_i^z)(1+\sigma_j^z)\\
    &+\frac{1}{8}\sum_{i\neq j}\Omega^{g,g}_{i,j}(1-\sigma_i^z)(1-\sigma_j^z)\\
    &+\frac{1}{4}\sum_{i\neq j}\Omega^{g,e}_{i,j}(1-\sigma_i^z)(1+\sigma_j^z).
  \end{split}
\end{equation}
This describes the unitary part of the interaction corresponding to a full-fledged spin-1/2 XXZ model. Here the detuning is defined as $\Delta=\omega-\omega_\ell$. The second term gives the usual electric and magnetic dipole-dipole interactions allowing for the hopping of excitations within the whole ensemble, describing the XY part. The last three terms represent the contribution from the static magnetic interactions and lead to frequency shifts conditioned on the occupancy of the pair of atoms involved in the interaction, which is typically referred to as an effective Ising interaction. The coherent photon exchange via electric dipole-dipole interactions $g^{(d)}_{ji}$ has been widely studied~\cite{Bromley2016}. The magnetic transition dipole-dipole couplings $g^{(\mu)}_{ji}$ generally can be ignored as they are of very small magnitude compared to the electric ones. However, other important couplings occur
\begin{equation}\label{eq:mag_interaction_form}
  \Omega^{\alpha,\beta}_{i,j}=\frac{\mu_0}{4\pi R_{ij}^3}\left[3\frac{(\bm\mu_{i,\alpha}\cdot\bm R_{ij})(\bm\mu_{j,\beta}\cdot\bm R_{ij})}{R_{ij}^2}-\bm\mu_{i,\alpha}\cdot\bm\mu_{j,\beta}\right]`
\end{equation}
with $\alpha,\beta\in\{e,g\}$, which describe density-density interactions between atoms owing to the static magnetic dipoles, illustrated in Fig.~\ref{fig2} on the right side.

With the elimination of the electromagnetic vacuum, the system is characterized by open system dynamics where the collective dissipation is included in the master equation
\begin{equation}
  \label{eq:mastereq}
  \dv{\rho}{t}=-\iu\commutator{\mathcal{H}_{\text{eff}}}{\rho}+\mathcal{L}_\mu[\rho]+\mathcal{L}_d[\rho].
\end{equation}
We assume standard Lindblad form for the loss terms which we write as follows
\begin{equation}\label{eq:dissipator}
  \mathcal{L}_\alpha[\rho]=\sum_{i,j=1}^\mathcal{N}f^{(\alpha)}_{ij}\left(2\sigma_i\rho\sigma_j^\dagger-\anticommutator{\sigma_j^\dagger\sigma_i}{\rho}\right).
\end{equation}
The independent radiative loss rates are $\Gamma=k_0^3d^2/(3\pi\hbar\epsilon_0)$ and $\Gamma_{\mu}=k_0^3\mu^2/3\pi\hbar\epsilon_0$ stemming from electric and magnetic contributions. Moreover, collective dissipation at rates $f^{(d)}_{ij}$ (electric) and $f^{(\mu)}_{ij}$ (magnetic) are also present, with exact expressions listed in App.~\ref{sec:elimination}.

\subsection{The weak-excitation limit: coupled dipoles model}
\label{sec:weak_excitation}

The Hamiltonian in Eq.~\eqref{eq:effective_hamiltonian} is of general validity. However, by considering the weak-excitation limit, the evolution of the system is restricted to a very small subspace where the analytical description of the dynamics can be greatly simplified. In order to do this, we linearize the time evolution of the system by replacing all population operators with $-1$ and factorizing all two operator correlations. We gather all expectation values of atomic coherences in a single vector $\bm v=(\expval{\sigma_1},...,\expval{\sigma_\mathcal{N}})^\top$ and write an effective first-order differential equation
\begin{equation}\label{eq:linear_pol_eqn}
    \dot{\bm v}(t)=-i \mathcal{M} \bm v (t)+\Omega \bm v_\text{drive},
\end{equation}
where the drive vector incorporates all the drive phases $\bm v_\text{drive}=(\me^{\iu{\bm k}_\ell \bm R_1},...,\me^{\iu {\bm k}_\ell\bm R_\mathcal{N}})^\top$.
The weak driving approximation $\Omega\ll\abs{\bm f^{(d)}}$ assumes that the Rabi frequency is much weaker than the dissipative part of the dipole-dipole interaction. Notice also that the equation above can be solved both in steady state to derive spectroscopic features of the ensemble as well as in the time domain (by imposing a time dependence on $\Omega$). The matrix $\mathcal{M}$ incorporates both coherent and dissipative cooperative behavior
\begin{equation}
  \mathcal{M}_{jj'}=\left[\Delta+\delta\omega_j\right]\delta_{jj'}-\left[g_{jj'}^{(d)}+\iu f_{jj'}^{(d)}\right].
\end{equation}
We used the low-excitation condition in order to approximate $\expval{\sigma^{z}_j\sigma_{j'}^{\phantom{\dagger}}}\approx -\expval{\sigma_{j'}^{\phantom{\dagger}}}$. This is always true if $j=j'$, but only approximately valid under the condition that very few excitations are present in the system such that, on average, each site has much lower than unit population in the excited state. This leads to the definition of a local frequency shift
\begin{equation}
  \begin{split}
      \delta\omega_j&=\sum_{j'\neq j}\left(\Omega_{j'j}^{e,g}-\Omega_{j'j}^{g,g}\right)\\
      &=\frac{(\mu_e-\mu_g)\mu_g\mu_0}{4\pi}\sum_{j'\neq j}\frac{3z_{ij}^2/R_{ij}^2-1}{R_{ij}^3},
  \end{split}
\end{equation}
where $\delta\omega_j$ denotes the total shift acquired owing to magnetic interactions. We have assumed that all magnetic dipole moments in the ground state are $\mu_g\hat e_z$ and all magnetic moments in the excited state are $\mu_e\hat e_z$. This leads to the definition of the magnetic interaction rate 
\begin{equation}
  \label{eq:C_mu_le}
  \Omega_\mu=\frac{\mu_0(\mu_e-\mu_g)\mu_g}{4\pi}.
\end{equation}
Additionally, we have neglected contributions from the magnetic transition dipole moments couplings as they are much smaller than the electrically mediated couplings. We then consider the coherent part of the far-field intensity radiated by a system defined by these dipoles
\begin{equation}
  \label{eq:Intensity}
  I(\bm r_s)\propto\left|\sum_{i}(1-z_s^2)\me^{-\iu\bm k_s\bm R_i}v_i\right|^2
\end{equation}
where $\bm r_s$ indicates the detection angle, $\bm k_s\parallel \bm r_s$ and $\abs{\bm k_s}=k_0$ is the wave vector for the propagation in detection direction.

\begin{figure}[t]
  \centering
  \includegraphics[width=\columnwidth]{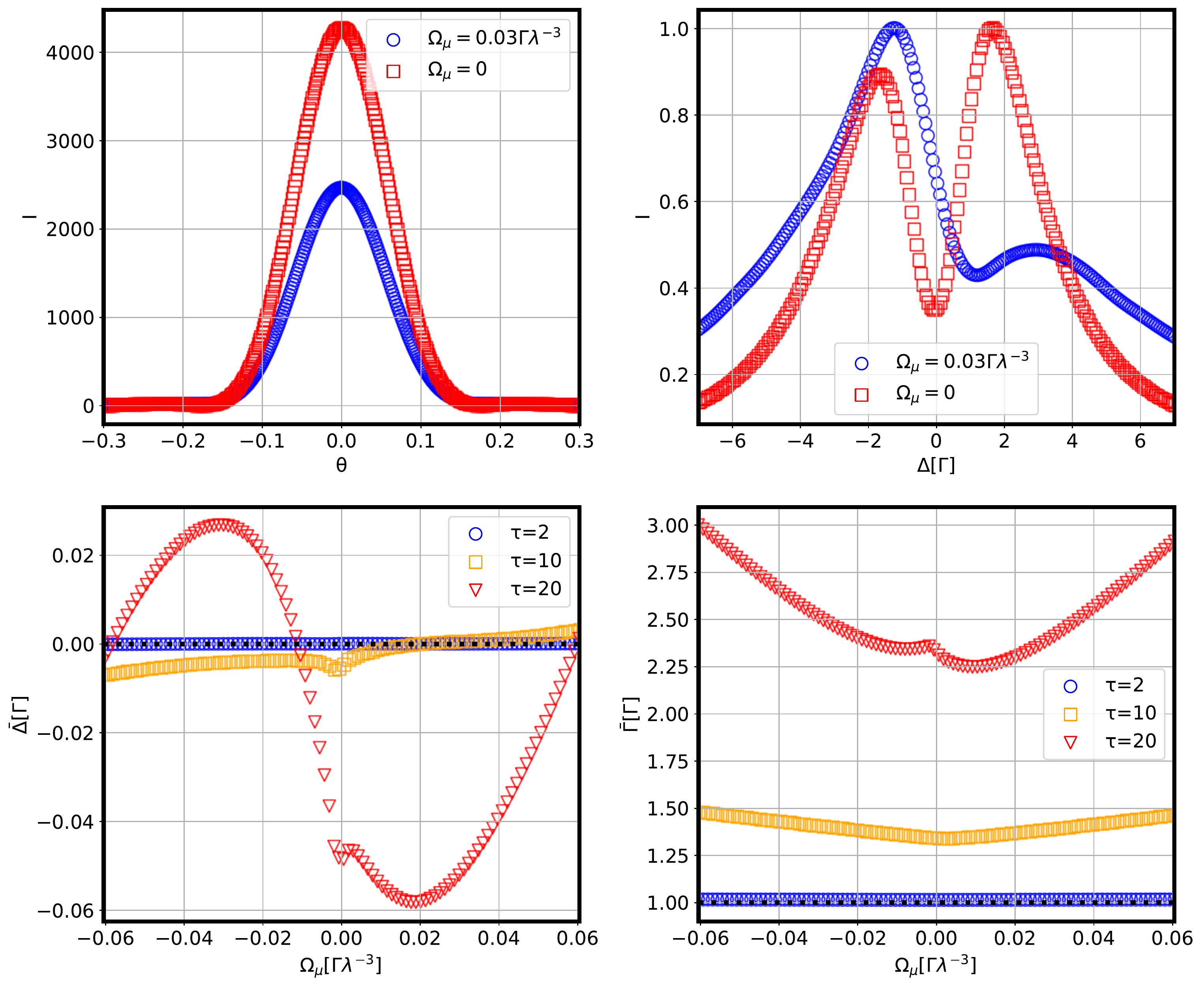}
  \caption{Numerical results based on solution of Eq.~\ref{eq:linear_pol_eqn} in steady state. We consider multiple configurations of a cloud where the particles are distributed according to a Gaussian probability distribution with optical depth $\tau=3\mathcal{N}/(2k_0^2R^2)$, simulating averaging over atomic motion. The number of configurations is $\mathcal {N}_c=1000$ and $\mathcal N=1000$. The optical depth is $\tau=15$. $\Omega_\mu$ quantifies the strength of the magnetic interaction. In (a) we show how the forward enhancement is modified by the magnetic interaction by plotting the intensity variation versus the in-plane angle $\theta$ on the single-particle resonance $\Delta=0$. There is a reduction in the forward enhancement for $\Omega_\mu=0.03\Gamma\lambda^{-3}$ ($\Omega_\mu k_0^{-3}=8\Gamma$) compared to no magnetic interaction. In (b) we investigate the lineshape of the cloud by changing the laser detuning $\Delta$ and compare $\Omega_\mu=0.03\Gamma\lambda^{-3}$ ($\Omega_\mu k_0^{-3}=8\Gamma$) with $\Omega_\mu=0$ at an in-plane detection angle of $\theta=\pi/10$. The double peak structure in the non-magnetic case becomes asymmetric when the magnetic interaction is turned on. Changing the sign flips the effect on the two peaks (not shown). In (c) and (d) we investigate the effect of the magnetic interaction on the cooperative properties of the sample. To this end we determine the collective lineshift $\bar\Delta$ and the collective linewidth $\bar\Gamma$ in the forward direction $\theta=0$ by calculating the lineshapes and then extracting these parameters from a Lorentzian fit. The lineshift and linewidth enhancement are shown for different optical densities $\tau$ with the magnetic interaction strength on the x-axis.}
  \label{fig3}
\end{figure}

In steady state, assuming time-independent driving, the linear system obtained from setting the left-hand side of Eq.\eqref{eq:linear_pol_eqn} to zero has to be solved. In practice, this equation needs to be solved for many different configurations. Since we assume the cloud to be thermal, we assume this particle distribution with respect to the center of the cloud to be Gaussian.
In Fig.~\ref{fig3}(a), we can see the straightforward effect of a reduction of the collective motion and consequently, the forward scattering enhancement due to frequency disorder. This is simply due to the fact that motional disorder increases the coupling between the driven mode and modes which do not emit in the laser direction.
The lineshape of the light scattered from a cloud, for two values of the magnetic interaction strength $\Omega_\mu$, is shown in Fig.~\ref{fig3}(b). For the dense ensemble we can already observe a double peak structure. The magnetic field shifts the spectral weight of the peaks to yield an asymmetric double peak structure which can be interpreted as a spectral shift of the eigenvalues due to the frequency disorder.

In Fig.~\ref{fig3}(c) and (d) we perform studies of the lineshift and linewidth modification due to the presence of the magnetic interaction for different optical densities $\tau$. From the lineshift we can estimate the total symmetric contribution to the frequency shift of the ensemble, which causes a lineshift that depends on the sign of the magnetic dipole-dipole interaction. From the linewidth enhancement one can see the effect of the asymmetric contribution to the frequency disorder. From $\tau=20$ it is fairly clear that there are two different regimes. For large magnetic interaction $\Omega_\mu$ the large frequency disorder simply leads to an inhomogeneous linewidth enhancement. For smaller $\Omega_\mu\approx 0.005\lambda^{-3}\Gamma$ ($\Omega_\mu k_0^{-3}\approx\Gamma_0$) the linewidth is actually decreased. One explanation for this is that the frequency disorder facilitates scattering into states which are naturally more subradiant than the originally driven state without disorder.

\section{Quantum degenerate case}
\label{sec4}
We consider now the case of a quantum degenerate gas and investigate the effect of the quantum statistical properties of the particles. As a first step, we follow a procedure which is standard for the derivation of the Dicke superradiance phenomenon with inverted two-level systems. This means, we will consider a non-interacting bosonic/fermionic trapped gas and will ask how does the emitted intensity scale as a function of the quantum statistics and as a function of trap parameters.

We start with a quantum degenerate gas in contact with a thermal reservoir at controllable temperature $T$. Tuning $T$ allows one to go from the quantum degenerate case to a completely classical, thermal gas regime (for high temperature). We will compute the intensity $I(\bm k_s)$ of angle-resolved spontaneous emission from the system with $\bm k_s$ being the wave vector in direction of the detector and $|\bm k_s|=k_0$. This intensity is defined as the correlation of the Fourier transforms of the polarization operators
\begin{equation}
  \label{eq:intensity}
  \begin{split}
  I(\bm k_s) \propto \expval{\hat \Sigma^\dagger(\bm k_s)\hat \Sigma(\bm k_s)},
\end{split}
\end{equation}
where the Fourier transformation is defined as follows $\hat \Sigma(\bm k_s)=\int\diff{\bm R}\me^{-\iu\bm k_s\bm R}\hat \Sigma (\bm R)$. Here $\expval{}$ stands for the ensemble average. We take the state in the following to be thermal in the excited and ground state so that the average becomes a thermal one where excited and ground state factorize.

For non-interacting scatterers in thermal equilibrium, the expectation value of a generic observable $\hat A$ can be written as
\begin{equation}
  \label{eq:thermal}
  \expval{\hat A}=\frac{1}{Z}\sum_{\bm N}\me^{-\beta \omega(\bm N)}\matrixelement{\bm N}{\hat A}{\bm N},
\end{equation}
where $\bm{N}=(\mathcal{N}_{g_0},\mathcal{N}_{e_0},\mathcal{N}_{g_1},\mathcal{N}_{e_1},\ldots)$ defines a microstate where the subindices $0,1,2,\ldots$ represent higher dimensional tuples of indices. This notation simply indicates that there are $\mathcal{N}_{\alpha_i}$ particles in the state with quantum numbers $\alpha_i$. Denoted by $\ket{\bm N}$ is the quantum state corresponding to these quantum numbers and $\omega(\bm N)$ is the total energy of the state. In the case of harmonically (frequency $\omega_t$) trapped atoms, the total energy  decomposes into summands
\begin{equation}
\omega(\bm N)=\omega_0 \sum_n \mathcal{N}_{e_n}+\omega_t\sum_n n\left(\mathcal{N}_{g_n}+\mathcal{N}_{e_n}\right).
\end{equation}
Let us first remark that, at moderate temperatures, owing to the fact that $\omega_0\ll \omega_t$, one can safely neglect the thermally induced occupancy of the excited electronic state, meaning that the first term containing $\omega_0$ can be ignored in the exponent $\me^{-\beta \omega(\bm N)}$.

In order to estimate the angle resolved scattered intensity we will make use of the trap basis. We utilize the decomposition of the field operators in this basis $\Psi_g(\bm R)=\sum_{\bm n}\phi_{\bm n}(\bm R)\hat g_{\bm n}$ and $\Psi_e(\bm R)=\sum_{\bm n}\phi_{\bm n}(\bm R)\hat e_{\bm n}$
and expand the two position correlation function as follows
\begin{align}
  \label{eq:sigma_representation}
    \expval{\hat \Sigma^\dagger(\bm R)\hat\Sigma(\bm R')}&=\sum_{\substack{\bm n_1,\bm n_2\\\bm m_1,\bm m_2}}\phi^*_{\bm m_1}(\bm R) \phi_{\bm n_1}(\bm R) \phi_{\bm m_2}(\bm R') \phi^*_{\bm n_2}(\bm R')\notag\\
    &\phantom{\sum_{\bm n_1,\bm n_2}}\cross\expval{\hat e_{\bm m_1}^\dagger\hat g_{\bm n_1}^{\phantom{\dagger}}\hat g_{\bm n_2}^\dagger\hat e_{\bm m_2}^{\phantom{\dagger}}}.
\end{align}
In order to compute the expression in Eq.~\eqref{eq:intensity} we apply the Fourier transform definition and identify the Franck-Condon factors stemming from the integration over space
\begin{equation}
  \label{eq:Franck_condon}
  \eta_{\bm n,\bm m}=\int\diff{\bm R}\me^{-\iu\bm R\bm k_s}\phi_{\bm n}^*(\bm R)\phi_{\bm m}^{\phantom{\dagger}}(\bm R).
\end{equation}
Next, we determine the thermal expectation values. In doing so, we ignore the contribution in the exponential stemming from $\omega_0$, corresponding to the energy of the electronic excited state, as its thermal occupancy is negligible even at room temperature. The result for the four-operator correlation becomes
\begin{equation}
  \label{eq:value_of_expectation_value}
  \expval{\hat e_{\bm m_1}^\dagger\hat g_{\bm n_1}^{\phantom{\dagger}}\hat g_{\bm n_2}^\dagger\hat e_{\bm m_2}}^{\phantom{\dagger}}=\delta_{\bm n_1\bm n_2}\delta_{\bm m_1\bm m_2}(\zeta p_{g_{\bm n}}+1)p_{e_{\bm m}},
\end{equation}
where $\sum_{\bm n} p_{g_{\bm n}}=\mathcal{N}_g$ and $\sum_{\bm n} p_{e_{\bm n}}=\mathcal{N}_e$ are the number of excitations in the ground and excited electronic bands, respectively, and $p_{g_{\bm n}}$ and $p_{e_{\bm n}}$ are the corresponding number distributions indexed by trap state $\bm n$. The parameter $\zeta$ is defined with two values: $\zeta=1$ for bosons and $\zeta=-1$ for fermions.
These results can be combined to find a general expression for the scaling of the radiated intensity
\begin{equation}
  \label{eq:intensity_analyt}
  I(\bm k_s)\propto \mathcal{N}_e+\zeta \sum_{\bm n,\bm m}\abs{\eta_{\bm n,\bm m}(\bm k_s)}^2p_{g_{\bm n}}p_{e_{\bm m}}.
\end{equation}
The first term in the expression of $I(\bm k_s)$ comes from independent emission, while the effect of quantum statistics and correlations comes from the second term $\sum_{n,m}\abs{\eta_{n,m}(\bm k_s)}^2p_{g_{\bm n}}p_{e_{\bm m}}$. Some insight can be obtained from limiting cases. The infinite-temperature case is trivial as the expected emission for either bosons or fermions is fully independent and is proportional to $\mathcal{N}_e$, as expected from a classical thermal gas, i.e. $I_{\text{bos/ferm}}(\bm k_s,T=\infty)\propto \mathcal{N}_e$ where all particles emit independently. The zero-temperature case is much more intriguing as we obtain the following expressions:
\begin{subequations}\label{eq:infinite_zero_temp}
  \begin{align}
    I_{\text{bos}}(\bm k_s,T=0)&\propto\me^{-\left(\bm k_s \cdot \bm r_\text{zpm}\right)^2}\mathcal{N}_e\left(\mathcal{N}_g+1\right)\label{eq:infinite_zero_temp_a},\\
    I_{\text{ferm}}(\bm k_s,T=0)&\propto \mathcal{N}_e-\sum_{\bm n,\bm m}^{\text{Filling}}\abs{\eta_{\bm n,\bm m}}^2\label{eq:infinite_zero_temp_c},
  \end{align}
\end{subequations}
for bosons and fermions respectively.

\begin{figure}[t]
  \includegraphics[width=0.89\columnwidth]{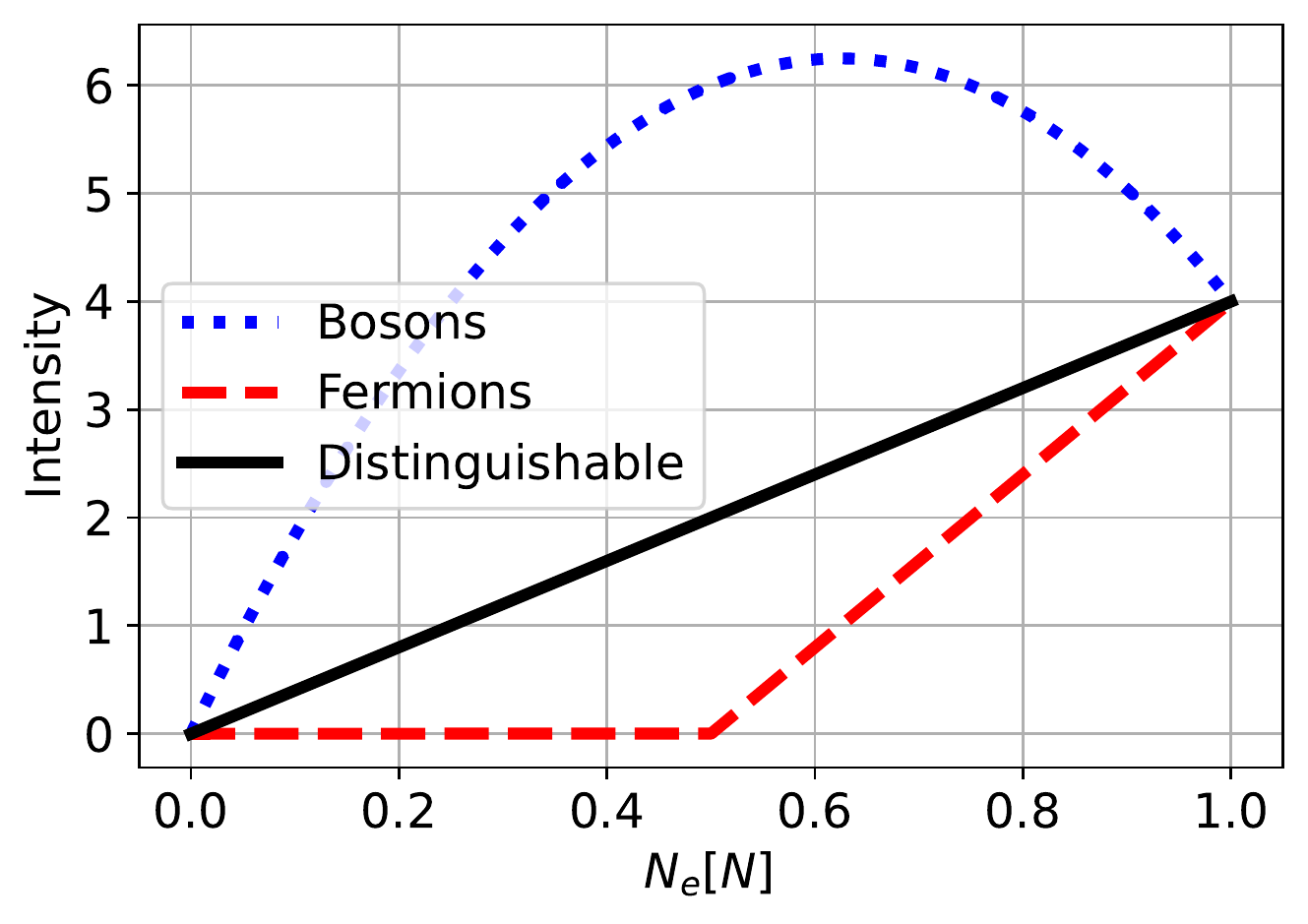}%
  \caption{Comparison between scattered intensity from distinguishable, fermionic and bosonic systems in the deep trap limit where the equations Eq.~\eqref{eq:infinite_zero_temp} apply, for $\mathcal{N}=4$. The intensity is shown as a function of the excited particle fraction $\mathcal{N}_e/\mathcal{N}$. The blue dotted line shows bosonic particles, the red dashed line fermionic particles and the solid black line distinguishable particles (which is the case for a high temperature ensemble).}
  \label{fig4}
\end{figure}

Let us first discuss the emission from a gas of bosons condensed onto the ground state of the external trap (which is the case at zero temperature). The Franck-Condon factor $\eta_{0,0}(\bm k_s)$ in this case is easily estimated and it includes the zero point motion $\bm r_\text{zpm}$  which indicates the extent of the wavefunction in the ground state of the harmonic trap. For deep trapping conditions, the Franck Condon factor is close to unity and the emission shows the same features as observed in standard Dicke superradiance in quantum optics. For sake of comparison, let us consider a fixed number of particles $\mathcal{N}=\mathcal{N}_e+\mathcal{N}_g$. For an ensemble of particles at the same spot, as assumed in the Dicke model, it is usually convenient to use the set of collective Dicke states $\ket{J,m}$, with $J=\mathcal{N}/2$ and $m=(\mathcal{N}_g-\mathcal{N}_e)/2$, situated on the surface of the Bloch sphere. Superradiance refers then to a situation where the emission of the collective state situated close to the equator shows a proportionality to $J^2$. By considering deeply trapped particles instead, the same kind of quick emission emerges as a fundamental property of the zero-temperature bosonic wavefunction. This is to be expected, as both models assume indistinguishable particles: in Dicke superradiance, this emerges from the condition that all emitters are placed in the same spot while for the bosonic cloud this is achieved by the symmetry of the ground state. The scaling of the emitted intensity with the number of excited atoms is seen in Fig.~\ref{fig4}.

In the fermionic case, the summation limit ``Filling'' in the equation above indicates that indices run from $(0,0)$ to the indices of the state at the Fermi energy level. The summation in Eq.~\eqref{eq:infinite_zero_temp_c} can be simplified in the case of a deep trap where $\eta_{\bm n,\bm m}\approx \delta_{\bm n,\bm m}$ so that $I_{\text{ferm}}(\bm k_s,T=0)\propto \mathcal{N}_e- \min(\mathcal{N}_e,\mathcal{N}_g)$. This can be understood as a dynamic suppression of spontaneous emission into occupied states due to the Pauli exclusion principle. For example, notice that below the equator of the Bloch sphere all states are fully subradiant. This is illustrated in Fig.~\ref{fig4} as the red, dashed curve and compared in Fig.\ref{fig4} against the case of distinguishable particles (also reached for high temperatures) where a linear scaling with the number of $\mathcal{N}_e$ is obtained.

\section{Experimental realization}
\label{sec5}

In order to experimentally observe the effects discussed in the previous sections, we have developed and built an apparatus to cool and trap neutral dysprosium \cite{Muhlbauer2018}.
Dysprosium belongs to the group of lanthanide elements whose characteristic open f-shell electron configuration [Xe]4f$^{10}$6s$^{2}$ with spin S = 2, orbital angular momentum L = 6, and total angular momentum J = 8 gives rise to its high magnetic moment of 10 Bohr magnetons ($\mu \sim 10 \mu_{B}$). Compared to alkali atoms, whose magnetic moments are on the order of only $\mu \sim 1 \mu_{B}$, the magnetic dipole-dipole interactions in ultra-cold gases of dysprosium is about 100 times stronger, making it an ideal candidate for investigating its contribution to cooperative effects. In addition, there is an almost equal abundance of stable bosonic and fermionic isotopes allowing for the creation of both types of quantum degenerate gases \cite{benlevbec2011,benlevfermi2012}.

The general experimental scheme for producing laser cooled samples of lanthanide atoms is implemented as follows \cite{ferlaino2012,pfau2014}: A strong optical transition in the blue spectrum range is used to precool the atoms in a Zeeman slower (ZS) before capturing them in a narrow line magneto optical trap (MOT). Operating a MOT on transitions with natural linewidths on the order of 100 kHz is required to reach Doppler temperatures below 5$\mu$K, which allows one to directly transfer the atoms from the MOT into an optical dipole trap (ODT). In our setup, we employ a strong J = 8 $\rightarrow$ J$^{\prime}$ = 9 transition at 421 nm, with a natural linewidth of $\Gamma_{421}= 2\pi\cdot32$ MHz for precooling. A thermal beam of atomic dysprosium, transversally cooled on this broad transition, reaches the ZS with initial velocities of several hundred meters per second. The atoms are then longitudinally decelerated to a velocity of about 24 m/s in the spin-flip configuration ZS before entering the main chamber, where they are captured in a 6 beam 3D MOT setup. The MOT transition, on the other hand, is a closed, narrow linewidth J = 8 $\rightarrow$ J$^{\prime}$ = 9 transition at 626 nm with a natural linewidth of $\Gamma_{626}= 2\pi\cdot136$ kHz which corresponds to a Doppler temperature of 3.2 $\mu$K.

\begin{figure}[t]
  \includegraphics[width=0.9\columnwidth]{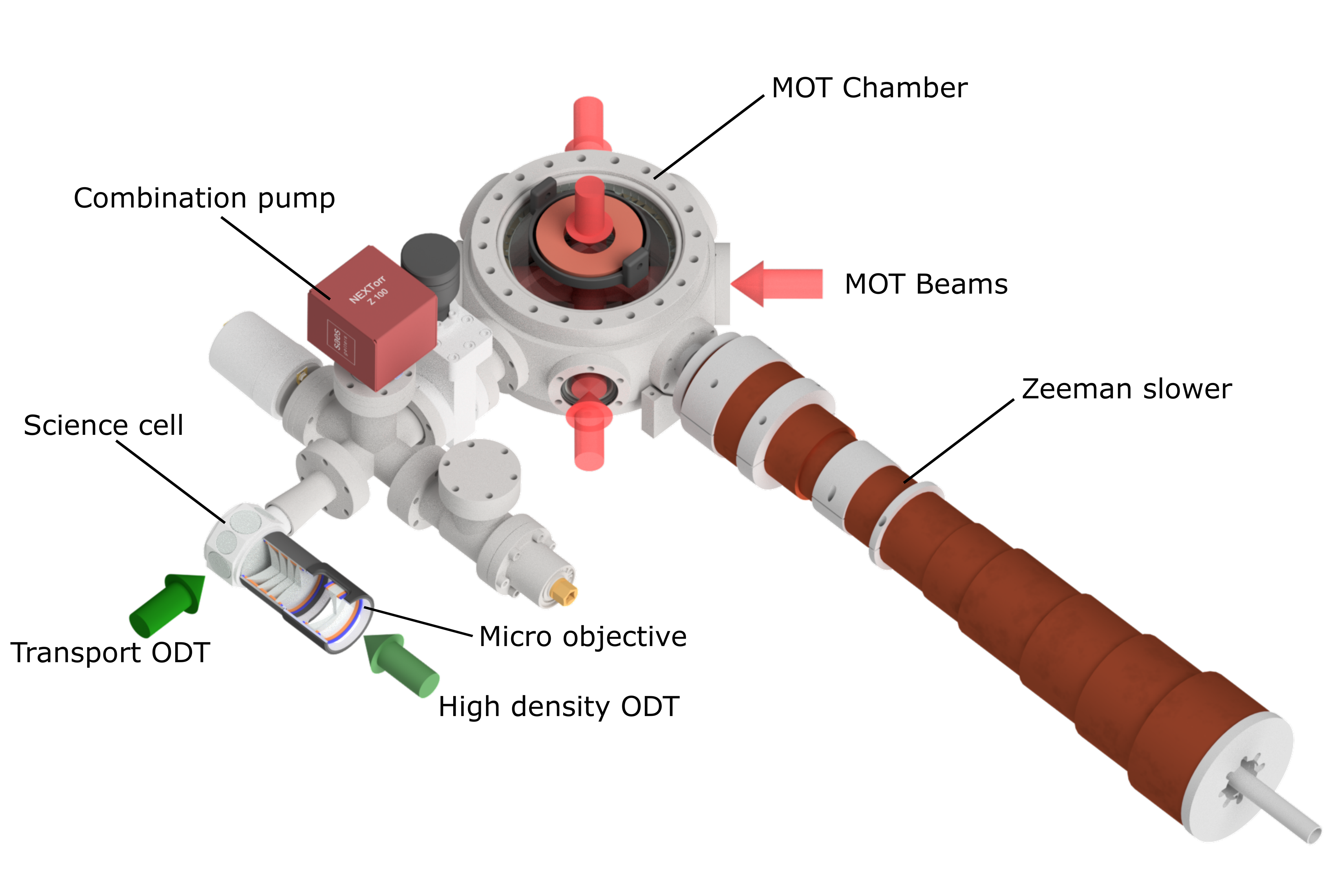}
  \caption{CAD render of the vacuum system. Dy atoms reach the ZS with a velocity of few hundred meters per second. They are longitudinally slowed in the ZS with resonant laser light at the 421 nm  before being captured in a narrow line 3D MOT at 626 nm. An achromatic lens in conjunction with an air-bearing translation stage is used to focus down the optical transport beam at the position of the MOT and create a deep ODT. The focal spot of this beam is then moved from the MOT chamber to the center of the science cell by moving the translation stage, thereby transferring the atoms alongside. In order to retain a substantial number of atoms in the trap after this sequence, the process of transport and re-trapping of atoms needs to be done on a timescale that is shorter than the lifetime of the ODT. }
  \label{fig5}
\end{figure}

Our experimental setup consists of a high numerical aperture (NA) science cell connected to an ultra-high vacuum (UHV) chamber (MOT) and a dedicated laser system capable of generating tunable and frequency stabilized laser radiation. Consequently, the laser frequency can be readily adapted to capture either bosons or fermions without any optical or mechanical adjustments. Fig.~\ref{fig5} shows a CAD render of the vacuum system. To reduce the influence of undesired external magnetic fields, the vacuum system is assembled exclusively from non-magnetic stainless steel, titanium and ceramic glass. Additionally, a commercial, three axis magnetic field compensation system (Stefan Mayer Instruments GmbH: MR-3) is used to drive a 3D coil system and compensate low frequency magnetic field disturbances.

The next step for creating dense samples of atomic dysprosium is to transfer atoms to the science cell attached to the MOT chamber. The cell offers a high optical access (with nine optical view ports), is designed in-house and made out of the machinable glass-ceramic MACOR (Corning Inc.). A deep ODT at the MOT position will be created, using a diode pumped solid-state (DPSS) laser (Coherent: Mephisto MOPA 55 W) in order to transport the atoms. This will be achieved by using an achromatic lens to focus down the transport beam to a beam waist of $\omega_{0} \sim 66 \mu m$ and two mirrors mounted on an air-bearing translation stage (Aerotech: ABL1500-300) which enables reproducible and precise movement of the focal spot of the beam, thereby, transporting the atoms alongside \cite{ketterletransport2001}. The atoms are re-trapped inside the science cell using a custom made multi-lens micro objective, designed to offer diffraction-limited performance at the trapping wavelength of 1064 nm\cite{Alt2001,Liobj2018}. A five-lens configuration consisting solely of commercial singlets was chosen to achieve a high NA of 0.53 and a working distance of 22 mm. Simulation and optimization of objective parameters like lens curvature, relative spacing, and thickness was done using a ray tracing software (Opticstudio) in order to minimize optical aberrations and achieve diffraction-limited performance. The beam waist at the focus of the micro-objective was measured using a piezo-controlled knife edge and found to be $\omega_{0h} = 5.94 \pm 1.18 \mu$m and $\omega_{0h} = 6.99 \pm 0.67 \mu$m, in the horizontal and vertical direction, respectively. We estimate that an ODT created with this objective should provide a sufficiently tight confinement for a few hundred to a few thousand dysprosium atoms to reach densities on the order of $\sim 10^{13}$ atoms/cm$^{3}$. This condition would allow then us to explore dynamics in the high-density regime where the inter-particle/inter-emitter spacing is lower than the wavelength of the exciting transition, giving rise to effects like Dicke sub- and superradiance.

\section{Conclusions}
\label{sec6}
We have considered the emission and scattering properties of a dense atomic cloud first by assuming distinguishable, classically moving particles and then moving onto the quantum degenerate case, where quantum statistics play an important role. In the case where atomic motion is treated classically, we made use of an open quantum system approach where we have shown that tunable spin Hamiltonians can be designed. In addition, under the assumption of weak external drive, and in particular for atoms where magnetic interactions are strong, we have characterized their effects onto the properties of scattered light. For quantum degenerate gases, in the non-interacting regime, we have quantified collective emission properties for both bosons and fermions. We expect that our findings are of direct relevance for dense atomic dysprosium clouds for which we describe an explicit experimental set up. In the future, we will explicitly include the interaction of atoms via the electromagnetic vacuum in order to derive an open system dynamics approach tailored to lossy quantum degenerate gases.

\section*{Acknowledgments}
We acknowledge financial support from the Max Planck Society and from the Deutsche Forschungsgemeinschaft (DFG, German Research Foundation) -- Project-ID 429529648 -- TRR 306 QuCoLiMa ("Quantum Cooperativity of Light and Matter’’).

\FloatBarrier
\bibliographystyle{apsrev4-1-custom}
\bibliography{bibliography}

\begin{thebibliography}{38}%
\makeatletter
\providecommand \@ifxundefined [1]{%
 \@ifx{#1\undefined}
}%
\providecommand \@ifnum [1]{%
 \ifnum #1\expandafter \@firstoftwo
 \else \expandafter \@secondoftwo
 \fi
}%
\providecommand \@ifx [1]{%
 \ifx #1\expandafter \@firstoftwo
 \else \expandafter \@secondoftwo
 \fi
}%
\providecommand \natexlab [1]{#1}%
\providecommand \enquote  [1]{``#1''}%
\providecommand \bibnamefont  [1]{#1}%
\providecommand \bibfnamefont [1]{#1}%
\providecommand \citenamefont [1]{#1}%
\providecommand \href@noop [0]{\@secondoftwo}%
\providecommand \href [0]{\begingroup \@sanitize@url \@href}%
\providecommand \@href[1]{\@@startlink{#1}\@@href}%
\providecommand \@@href[1]{\endgroup#1\@@endlink}%
\providecommand \@sanitize@url [0]{\catcode `\\12\catcode `\$12\catcode
  `\&12\catcode `\#12\catcode `\^12\catcode `\_12\catcode `\%12\relax}%
\providecommand \@@startlink[1]{}%
\providecommand \@@endlink[0]{}%
\providecommand \url  [0]{\begingroup\@sanitize@url \@url }%
\providecommand \@url [1]{\endgroup\@href {#1}{\urlprefix }}%
\providecommand \urlprefix  [0]{URL }%
\providecommand \Eprint [0]{\href }%
\providecommand \doibase [0]{http://dx.doi.org/}%
\providecommand \selectlanguage [0]{\@gobble}%
\providecommand \bibinfo  [0]{\@secondoftwo}%
\providecommand \bibfield  [0]{\@secondoftwo}%
\providecommand \translation [1]{[#1]}%
\providecommand \BibitemOpen [0]{}%
\providecommand \bibitemStop [0]{}%
\providecommand \bibitemNoStop [0]{.\EOS\space}%
\providecommand \EOS [0]{\spacefactor3000\relax}%
\providecommand \BibitemShut  [1]{\csname bibitem#1\endcsname}%
\let\auto@bib@innerbib\@empty
\bibitem [{\citenamefont {Labeyrie} \emph {et~al.}(1999)\citenamefont
  {Labeyrie}, \citenamefont {de~Tomasi}, \citenamefont {Bernard}, \citenamefont
  {M\"uller}, \citenamefont {Miniatura}, and \citenamefont
  {Kaiser}}]{Labeyrie99}%
  \BibitemOpen
  \bibfield  {author} {\bibinfo {author} {\bibfnamefont {G.}~\bibnamefont
  {Labeyrie}}, \bibinfo {author} {\bibfnamefont {F.}~\bibnamefont {de~Tomasi}},
  \bibinfo {author} {\bibfnamefont {J.-C.} \bibnamefont {Bernard}}, \bibinfo
  {author} {\bibfnamefont {C.~A.} \bibnamefont {M\"uller}}, \bibinfo {author}
  {\bibfnamefont {C.}~\bibnamefont {Miniatura}},  and \bibinfo {author}
  {\bibfnamefont {R.}~\bibnamefont {Kaiser}}, }\bibfield  {title} {\enquote
  {\bibinfo {title} {Coherent backscattering of light by cold atoms},} }\href
  {\doibase 10.1103/PhysRevLett.83.5266} {\bibfield  {journal} {\bibinfo
  {journal} {Phys. Rev. Lett.} }\textbf {\bibinfo {volume} {83}}, \bibinfo
  {pages} {5266--5269} (\bibinfo {year} {1999})}\BibitemShut {NoStop}%
\bibitem [{\citenamefont {Labeyrie} \emph {et~al.}(2003)\citenamefont
  {Labeyrie}, \citenamefont {Delande}, \citenamefont {M\"uller}, \citenamefont
  {Miniatura}, and \citenamefont {Kaiser}}]{Labeyrie03}%
  \BibitemOpen
  \bibfield  {author} {\bibinfo {author} {\bibfnamefont {G.}~\bibnamefont
  {Labeyrie}}, \bibinfo {author} {\bibfnamefont {D.}~\bibnamefont {Delande}},
  \bibinfo {author} {\bibfnamefont {C.~A.} \bibnamefont {M\"uller}}, \bibinfo
  {author} {\bibfnamefont {C.}~\bibnamefont {Miniatura}},  and \bibinfo
  {author} {\bibfnamefont {R.}~\bibnamefont {Kaiser}}, }\bibfield  {title}
  {\enquote {\bibinfo {title} {Coherent backscattering of light by an
  inhomogeneous cloud of cold atoms},} }\href {\doibase
  10.1103/PhysRevA.67.033814} {\bibfield  {journal} {\bibinfo  {journal} {Phys.
  Rev. A} }\textbf {\bibinfo {volume} {67}}, \bibinfo {pages} {033814}
  (\bibinfo {year} {2003})}\BibitemShut {NoStop}%
\bibitem [{\citenamefont {Dicke}(1954)}]{Dicke54}%
  \BibitemOpen
  \bibfield  {author} {\bibinfo {author} {\bibfnamefont {R.~H.} \bibnamefont
  {Dicke}}, }\bibfield  {title} {\enquote {\bibinfo {title} {Coherence in
  spontaneous radiation processes},} }\href {\doibase 10.1103/PhysRev.93.99}
  {\bibfield  {journal} {\bibinfo  {journal} {Phys. Rev.} }\textbf {\bibinfo
  {volume} {93}}, \bibinfo {pages} {99--110} (\bibinfo {year}
  {1954})}\BibitemShut {NoStop}%
\bibitem [{\citenamefont {Kwong} \emph {et~al.}(2015)\citenamefont {Kwong},
  \citenamefont {Yang}, \citenamefont {Delande}, \citenamefont {Pierrat}, and
  \citenamefont {Wilkowski}}]{Kwong2015}%
  \BibitemOpen
  \bibfield  {author} {\bibinfo {author} {\bibfnamefont {C.~C.} \bibnamefont
  {Kwong}}, \bibinfo {author} {\bibfnamefont {T.}~\bibnamefont {Yang}},
  \bibinfo {author} {\bibfnamefont {D.}~\bibnamefont {Delande}}, \bibinfo
  {author} {\bibfnamefont {R.}~\bibnamefont {Pierrat}},  and \bibinfo {author}
  {\bibfnamefont {D.}~\bibnamefont {Wilkowski}}, }\bibfield  {title} {\enquote
  {\bibinfo {title} {Cooperative emission of a pulse train in an optically
  thick scattering medium},} }\href {\doibase 10.1103/PhysRevLett.115.223601}
  {\bibfield  {journal} {\bibinfo  {journal} {Phys. Rev. Lett.} }\textbf
  {\bibinfo {volume} {115}}, \bibinfo {pages} {223601} (\bibinfo {year}
  {2015})}\BibitemShut {NoStop}%
\bibitem [{\citenamefont {Bromley} \emph {et~al.}(2016)\citenamefont {Bromley},
  \citenamefont {Zhu}, \citenamefont {Bishof}, \citenamefont {Zhang},
  \citenamefont {Bothwell}, \citenamefont {Schachenmayer}, \citenamefont
  {Nicholson}, \citenamefont {Kaiser}, \citenamefont {Yelin}, \citenamefont
  {Lukin}, \citenamefont {Rey}, and \citenamefont {Ye}}]{Bromley2016}%
  \BibitemOpen
  \bibfield  {author} {\bibinfo {author} {\bibfnamefont {S.~L.} \bibnamefont
  {Bromley}}, \bibinfo {author} {\bibfnamefont {B.}~\bibnamefont {Zhu}},
  \bibinfo {author} {\bibfnamefont {M.}~\bibnamefont {Bishof}}, \bibinfo
  {author} {\bibfnamefont {X.}~\bibnamefont {Zhang}}, \bibinfo {author}
  {\bibfnamefont {T.}~\bibnamefont {Bothwell}}, \bibinfo {author}
  {\bibfnamefont {J.}~\bibnamefont {Schachenmayer}}, \bibinfo {author}
  {\bibfnamefont {T.~L.} \bibnamefont {Nicholson}}, \bibinfo {author}
  {\bibfnamefont {R.}~\bibnamefont {Kaiser}}, \bibinfo {author} {\bibfnamefont
  {S.~F.} \bibnamefont {Yelin}}, \bibinfo {author} {\bibfnamefont {M.~D.}
  \bibnamefont {Lukin}}, \bibinfo {author} {\bibfnamefont {A.~M.} \bibnamefont
  {Rey}},  and \bibinfo {author} {\bibfnamefont {J.}~\bibnamefont {Ye}},
  }\bibfield  {title} {\enquote {\bibinfo {title} {Collective atomic scattering
  and motional effects in a dense coherent medium},} }\href {\doibase
  10.1038/ncomms11039} {\bibfield  {journal} {\bibinfo  {journal} {Nature
  Communications} }\textbf {\bibinfo {volume} {7}}, \bibinfo {pages} {11039}
  (\bibinfo {year} {2016})}\BibitemShut {NoStop}%
\bibitem [{\citenamefont {Inouye} \emph {et~al.}(1999)\citenamefont {Inouye},
  \citenamefont {Chikkatur}, \citenamefont {Stamper-Kurn}, \citenamefont
  {Stenger}, \citenamefont {Pritchard}, and \citenamefont
  {Ketterle}}]{Inouye99}%
  \BibitemOpen
  \bibfield  {author} {\bibinfo {author} {\bibfnamefont {S.}~\bibnamefont
  {Inouye}}, \bibinfo {author} {\bibfnamefont {A.~P.} \bibnamefont
  {Chikkatur}}, \bibinfo {author} {\bibfnamefont {D.~M.} \bibnamefont
  {Stamper-Kurn}}, \bibinfo {author} {\bibfnamefont {J.}~\bibnamefont
  {Stenger}}, \bibinfo {author} {\bibfnamefont {D.~E.} \bibnamefont
  {Pritchard}},  and \bibinfo {author} {\bibfnamefont {W.}~\bibnamefont
  {Ketterle}}, }\bibfield  {title} {\enquote {\bibinfo {title} {Superradiant
  rayleigh scattering from a bose-einstein condensate},} }\href {\doibase
  10.1126/science.285.5427.571} {\bibfield  {journal} {\bibinfo  {journal}
  {Science} }\textbf {\bibinfo {volume} {285}}, \bibinfo {pages} {571--574}
  (\bibinfo {year} {1999})}\BibitemShut {NoStop}%
\bibitem [{\citenamefont {Weiss} \emph {et~al.}(2019)\citenamefont {Weiss},
  \citenamefont {Cipris}, \citenamefont {Ara\'ujo}, \citenamefont {Kaiser}, and
  \citenamefont {Guerin}}]{weiss19}%
  \BibitemOpen
  \bibfield  {author} {\bibinfo {author} {\bibfnamefont {P.}~\bibnamefont
  {Weiss}}, \bibinfo {author} {\bibfnamefont {A.}~\bibnamefont {Cipris}},
  \bibinfo {author} {\bibfnamefont {M.~O.} \bibnamefont {Ara\'ujo}}, \bibinfo
  {author} {\bibfnamefont {R.}~\bibnamefont {Kaiser}},  and \bibinfo {author}
  {\bibfnamefont {W.}~\bibnamefont {Guerin}}, }\bibfield  {title} {\enquote
  {\bibinfo {title} {Robustness of dicke subradiance against thermal
  decoherence},} }\href {\doibase 10.1103/PhysRevA.100.033833} {\bibfield
  {journal} {\bibinfo  {journal} {Phys. Rev. A} }\textbf {\bibinfo {volume}
  {100}}, \bibinfo {pages} {033833} (\bibinfo {year} {2019})}\BibitemShut
  {NoStop}%
\bibitem [{\citenamefont {Jennewein} \emph {et~al.}(2018)\citenamefont
  {Jennewein}, \citenamefont {Brossard}, \citenamefont {Sortais}, \citenamefont
  {Browaeys}, \citenamefont {Cheinet}, \citenamefont {Robert}, and
  \citenamefont {Pillet}}]{jennewein18}%
  \BibitemOpen
  \bibfield  {author} {\bibinfo {author} {\bibfnamefont {S.}~\bibnamefont
  {Jennewein}}, \bibinfo {author} {\bibfnamefont {L.}~\bibnamefont {Brossard}},
  \bibinfo {author} {\bibfnamefont {Y.~R.~P.} \bibnamefont {Sortais}}, \bibinfo
  {author} {\bibfnamefont {A.}~\bibnamefont {Browaeys}}, \bibinfo {author}
  {\bibfnamefont {P.}~\bibnamefont {Cheinet}}, \bibinfo {author} {\bibfnamefont
  {J.}~\bibnamefont {Robert}},  and \bibinfo {author} {\bibfnamefont
  {P.}~\bibnamefont {Pillet}}, }\bibfield  {title} {\enquote {\bibinfo {title}
  {Coherent scattering of near-resonant light by a dense, microscopic cloud of
  cold two-level atoms: Experiment versus theory},} }\href {\doibase
  10.1103/PhysRevA.97.053816} {\bibfield  {journal} {\bibinfo  {journal} {Phys.
  Rev. A} }\textbf {\bibinfo {volume} {97}}, \bibinfo {pages} {053816}
  (\bibinfo {year} {2018})}\BibitemShut {NoStop}%
\bibitem [{\citenamefont {de~Oliveira} \emph {et~al.}(2014)\citenamefont
  {de~Oliveira}, \citenamefont {Mendes}, \citenamefont {Martins}, \citenamefont
  {Saldanha}, \citenamefont {Tabosa}, and \citenamefont
  {Felinto}}]{deoliviera14}%
  \BibitemOpen
  \bibfield  {author} {\bibinfo {author} {\bibfnamefont {R.~A.} \bibnamefont
  {de~Oliveira}}, \bibinfo {author} {\bibfnamefont {M.~S.} \bibnamefont
  {Mendes}}, \bibinfo {author} {\bibfnamefont {W.~S.} \bibnamefont {Martins}},
  \bibinfo {author} {\bibfnamefont {P.~L.} \bibnamefont {Saldanha}}, \bibinfo
  {author} {\bibfnamefont {J.~W.~R.} \bibnamefont {Tabosa}},  and \bibinfo
  {author} {\bibfnamefont {D.}~\bibnamefont {Felinto}}, }\bibfield  {title}
  {\enquote {\bibinfo {title} {Single-photon superradiance in cold atoms},}
  }\href {\doibase 10.1103/PhysRevA.90.023848} {\bibfield  {journal} {\bibinfo
  {journal} {Phys. Rev. A} }\textbf {\bibinfo {volume} {90}}, \bibinfo {pages}
  {023848} (\bibinfo {year} {2014})}\BibitemShut {NoStop}%
\bibitem [{\citenamefont {Roof} \emph {et~al.}(2016)\citenamefont {Roof},
  \citenamefont {Kemp}, \citenamefont {Havey}, and \citenamefont
  {Sokolov}}]{roof16}%
  \BibitemOpen
  \bibfield  {author} {\bibinfo {author} {\bibfnamefont {S.~J.} \bibnamefont
  {Roof}}, \bibinfo {author} {\bibfnamefont {K.~J.} \bibnamefont {Kemp}},
  \bibinfo {author} {\bibfnamefont {M.~D.} \bibnamefont {Havey}},  and \bibinfo
  {author} {\bibfnamefont {I.~M.} \bibnamefont {Sokolov}}, }\bibfield  {title}
  {\enquote {\bibinfo {title} {Observation of single-photon superradiance and
  the cooperative lamb shift in an extended sample of cold atoms},} }\href
  {\doibase 10.1103/PhysRevLett.117.073003} {\bibfield  {journal} {\bibinfo
  {journal} {Phys. Rev. Lett.} }\textbf {\bibinfo {volume} {117}}, \bibinfo
  {pages} {073003} (\bibinfo {year} {2016})}\BibitemShut {NoStop}%
\bibitem [{\citenamefont {Ara\'ujo} \emph {et~al.}(2016)\citenamefont
  {Ara\'ujo}, \citenamefont {Kre\ifmmode \check{s}\else
  \v{s}\fi{}i\ifmmode~\acute{c}\else \'{c}\fi{}}, \citenamefont {Kaiser}, and
  \citenamefont {Guerin}}]{kaiser16}%
  \BibitemOpen
  \bibfield  {author} {\bibinfo {author} {\bibfnamefont {M.~O.} \bibnamefont
  {Ara\'ujo}}, \bibinfo {author} {\bibfnamefont {I.}~\bibnamefont {Kre\ifmmode
  \check{s}\else \v{s}\fi{}i\ifmmode~\acute{c}\else \'{c}\fi{}}}, \bibinfo
  {author} {\bibfnamefont {R.}~\bibnamefont {Kaiser}},  and \bibinfo {author}
  {\bibfnamefont {W.}~\bibnamefont {Guerin}}, }\bibfield  {title} {\enquote
  {\bibinfo {title} {Superradiance in a large and dilute cloud of cold atoms in
  the linear-optics regime},} }\href {\doibase 10.1103/PhysRevLett.117.073002}
  {\bibfield  {journal} {\bibinfo  {journal} {Phys. Rev. Lett.} }\textbf
  {\bibinfo {volume} {117}}, \bibinfo {pages} {073002} (\bibinfo {year}
  {2016})}\BibitemShut {NoStop}%
\bibitem [{\citenamefont {Guerin} \emph {et~al.}(2016)\citenamefont {Guerin},
  \citenamefont {Ara\'ujo}, and \citenamefont {Kaiser}}]{guerin16}%
  \BibitemOpen
  \bibfield  {author} {\bibinfo {author} {\bibfnamefont {W.}~\bibnamefont
  {Guerin}}, \bibinfo {author} {\bibfnamefont {M.~O.} \bibnamefont {Ara\'ujo}},
   and \bibinfo {author} {\bibfnamefont {R.}~\bibnamefont {Kaiser}}, }\bibfield
   {title} {\enquote {\bibinfo {title} {Subradiance in a large cloud of cold
  atoms},} }\href {\doibase 10.1103/PhysRevLett.116.083601} {\bibfield
  {journal} {\bibinfo  {journal} {Phys. Rev. Lett.} }\textbf {\bibinfo {volume}
  {116}}, \bibinfo {pages} {083601} (\bibinfo {year} {2016})}\BibitemShut
  {NoStop}%
\bibitem [{\citenamefont {Gross} and \citenamefont {Haroche}(1982)}]{Gross82}%
  \BibitemOpen
  \bibfield  {author} {\bibinfo {author} {\bibfnamefont {M.}~\bibnamefont
  {Gross}} and \bibinfo {author} {\bibfnamefont {S.}~\bibnamefont {Haroche}},
  }\bibfield  {title} {\enquote {\bibinfo {title} {Superradiance: An essay on
  the theory of collective spontaneous emission},} }\href {\doibase
  https://doi.org/10.1016/0370-1573(82)90102-8} {\bibfield  {journal} {\bibinfo
   {journal} {Physics Reports} }\textbf {\bibinfo {volume} {93}}, \bibinfo
  {pages} {301--396} (\bibinfo {year} {1982})}\BibitemShut {NoStop}%
\bibitem [{\citenamefont {Scully} \emph {et~al.}(2006)\citenamefont {Scully},
  \citenamefont {Fry}, \citenamefont {Ooi}, and \citenamefont
  {W\'odkiewicz}}]{Scully04}%
  \BibitemOpen
  \bibfield  {author} {\bibinfo {author} {\bibfnamefont {M.~O.} \bibnamefont
  {Scully}}, \bibinfo {author} {\bibfnamefont {E.~S.} \bibnamefont {Fry}},
  \bibinfo {author} {\bibfnamefont {C.~H.~R.} \bibnamefont {Ooi}},  and
  \bibinfo {author} {\bibfnamefont {K.}~\bibnamefont {W\'odkiewicz}},
  }\bibfield  {title} {\enquote {\bibinfo {title} {Directed spontaneous
  emission from an extended ensemble of $n$ atoms: Timing is everything},}
  }\href {\doibase 10.1103/PhysRevLett.96.010501} {\bibfield  {journal}
  {\bibinfo  {journal} {Phys. Rev. Lett.} }\textbf {\bibinfo {volume} {96}},
  \bibinfo {pages} {010501} (\bibinfo {year} {2006})}\BibitemShut {NoStop}%
\bibitem [{\citenamefont {A.} \emph {et~al.}(2010)\citenamefont {A.},
  \citenamefont {Chang}, and \citenamefont {Scully}}]{Svidzinsky10}%
  \BibitemOpen
  \bibfield  {author} {\bibinfo {author} {\bibfnamefont {S.~A.} \bibnamefont
  {A.}}, \bibinfo {author} {\bibfnamefont {J.-T.} \bibnamefont {Chang}},  and
  \bibinfo {author} {\bibfnamefont {M.~O.} \bibnamefont {Scully}}, }\bibfield
  {title} {\enquote {\bibinfo {title} {Cooperative spontaneous emission of $n$
  atoms: Many-body eigenstates, the effect of virtual lamb shift processes, and
  analogy with radiation of $n$ classical oscillators},} }\href {\doibase
  10.1103/PhysRevA.81.053821} {\bibfield  {journal} {\bibinfo  {journal} {Phys.
  Rev. A} }\textbf {\bibinfo {volume} {81}}, \bibinfo {pages} {053821}
  (\bibinfo {year} {2010})}\BibitemShut {NoStop}%
\bibitem [{\citenamefont {Keaveney} \emph {et~al.}(2012)\citenamefont
  {Keaveney}, \citenamefont {Sargsyan}, \citenamefont {Krohn}, \citenamefont
  {Hughes}, \citenamefont {Sarkisyan}, and \citenamefont {Adams}}]{Keaveney12}%
  \BibitemOpen
  \bibfield  {author} {\bibinfo {author} {\bibfnamefont {J.}~\bibnamefont
  {Keaveney}}, \bibinfo {author} {\bibfnamefont {A.}~\bibnamefont {Sargsyan}},
  \bibinfo {author} {\bibfnamefont {U.}~\bibnamefont {Krohn}}, \bibinfo
  {author} {\bibfnamefont {I.~G.} \bibnamefont {Hughes}}, \bibinfo {author}
  {\bibfnamefont {D.}~\bibnamefont {Sarkisyan}},  and \bibinfo {author}
  {\bibfnamefont {C.~S.} \bibnamefont {Adams}}, }\bibfield  {title} {\enquote
  {\bibinfo {title} {Cooperative lamb shift in an atomic vapor layer of
  nanometer thickness},} }\href {\doibase 10.1103/PhysRevLett.108.173601}
  {\bibfield  {journal} {\bibinfo  {journal} {Phys. Rev. Lett.} }\textbf
  {\bibinfo {volume} {108}}, \bibinfo {pages} {173601} (\bibinfo {year}
  {2012})}\BibitemShut {NoStop}%
\bibitem [{\citenamefont {Javanainen} \emph {et~al.}(2014)\citenamefont
  {Javanainen}, \citenamefont {Ruostekoski}, \citenamefont {Li}, and
  \citenamefont {Yoo}}]{Javanainen14}%
  \BibitemOpen
  \bibfield  {author} {\bibinfo {author} {\bibfnamefont {J.}~\bibnamefont
  {Javanainen}}, \bibinfo {author} {\bibfnamefont {J.}~\bibnamefont
  {Ruostekoski}}, \bibinfo {author} {\bibfnamefont {Y.}~\bibnamefont {Li}},
  and \bibinfo {author} {\bibfnamefont {S.-M.} \bibnamefont {Yoo}}, }\bibfield
  {title} {\enquote {\bibinfo {title} {Shifts of a resonance line in a dense
  atomic sample},} }\href {\doibase 10.1103/PhysRevLett.112.113603} {\bibfield
  {journal} {\bibinfo  {journal} {Phys. Rev. Lett.} }\textbf {\bibinfo {volume}
  {112}}, \bibinfo {pages} {113603} (\bibinfo {year} {2014})}\BibitemShut
  {NoStop}%
\bibitem [{\citenamefont {Kaiser}(2009)}]{Kaiser09}%
  \BibitemOpen
  \bibfield  {author} {\bibinfo {author} {\bibfnamefont {R.}~\bibnamefont
  {Kaiser}}, }\bibfield  {title} {\enquote {\bibinfo {title} {Quantum multiple
  scattering},} }\href {\doibase 10.1080/09500340903082663} {\bibfield
  {journal} {\bibinfo  {journal} {Journal of Modern Optics} }\textbf {\bibinfo
  {volume} {56}}, \bibinfo {pages} {2082--2088} (\bibinfo {year} {2009})},
  \Eprint {http://arxiv.org/abs/https://doi.org/10.1080/09500340903082663}
  {https://doi.org/10.1080/09500340903082663} \BibitemShut {NoStop}%
\bibitem [{\citenamefont {Zhu} \emph {et~al.}(2016)\citenamefont {Zhu},
  \citenamefont {Cooper}, \citenamefont {Ye}, and \citenamefont
  {Rey}}]{Zhu2016}%
  \BibitemOpen
  \bibfield  {author} {\bibinfo {author} {\bibfnamefont {B.}~\bibnamefont
  {Zhu}}, \bibinfo {author} {\bibfnamefont {J.}~\bibnamefont {Cooper}},
  \bibinfo {author} {\bibfnamefont {J.}~\bibnamefont {Ye}},  and \bibinfo
  {author} {\bibfnamefont {A.~M.} \bibnamefont {Rey}}, }\bibfield  {title}
  {\enquote {\bibinfo {title} {Light scattering from dense cold atomic media},}
  }\href {\doibase 10.1103/PhysRevA.94.023612} {\bibfield  {journal} {\bibinfo
  {journal} {Phys. Rev. A} }\textbf {\bibinfo {volume} {94}}, \bibinfo {pages}
  {023612} (\bibinfo {year} {2016})}\BibitemShut {NoStop}%
\bibitem [{\citenamefont {Javanainen} and \citenamefont
  {Ruostekoski}(2016)}]{Javanainen2016}%
  \BibitemOpen
  \bibfield  {author} {\bibinfo {author} {\bibfnamefont {J.}~\bibnamefont
  {Javanainen}} and \bibinfo {author} {\bibfnamefont {J.}~\bibnamefont
  {Ruostekoski}}, }\bibfield  {title} {\enquote {\bibinfo {title} {Light
  propagation beyond the mean-field theory of standard optics},} }\href
  {\doibase 10.1364/OE.24.000993} {\bibfield  {journal} {\bibinfo  {journal}
  {Opt. Express} }\textbf {\bibinfo {volume} {24}}, \bibinfo {pages}
  {993--1001} (\bibinfo {year} {2016})}\BibitemShut {NoStop}%
\bibitem [{\citenamefont {Pellegrino} \emph {et~al.}(2014)\citenamefont
  {Pellegrino}, \citenamefont {Bourgain}, \citenamefont {Jennewein},
  \citenamefont {Sortais}, \citenamefont {Browaeys}, \citenamefont {Jenkins},
  and \citenamefont {Ruostekoski}}]{Pellegrino2014}%
  \BibitemOpen
  \bibfield  {author} {\bibinfo {author} {\bibfnamefont {J.}~\bibnamefont
  {Pellegrino}}, \bibinfo {author} {\bibfnamefont {R.}~\bibnamefont
  {Bourgain}}, \bibinfo {author} {\bibfnamefont {S.}~\bibnamefont {Jennewein}},
  \bibinfo {author} {\bibfnamefont {Y.~R.~P.} \bibnamefont {Sortais}}, \bibinfo
  {author} {\bibfnamefont {A.}~\bibnamefont {Browaeys}}, \bibinfo {author}
  {\bibfnamefont {S.~D.} \bibnamefont {Jenkins}},  and \bibinfo {author}
  {\bibfnamefont {J.}~\bibnamefont {Ruostekoski}}, }\bibfield  {title}
  {\enquote {\bibinfo {title} {Observation of suppression of light scattering
  induced by dipole-dipole interactions in a cold-atom ensemble},} }\href
  {\doibase 10.1103/PhysRevLett.113.133602} {\bibfield  {journal} {\bibinfo
  {journal} {Phys. Rev. Lett.} }\textbf {\bibinfo {volume} {113}}, \bibinfo
  {pages} {133602} (\bibinfo {year} {2014})}\BibitemShut {NoStop}%
\bibitem [{\citenamefont {Jenkins} \emph {et~al.}(2016)\citenamefont {Jenkins},
  \citenamefont {Ruostekoski}, \citenamefont {Javanainen}, \citenamefont
  {Bourgain}, \citenamefont {Jennewein}, \citenamefont {Sortais}, and
  \citenamefont {Browaeys}}]{Jenkins2016}%
  \BibitemOpen
  \bibfield  {author} {\bibinfo {author} {\bibfnamefont {S.~D.} \bibnamefont
  {Jenkins}}, \bibinfo {author} {\bibfnamefont {J.}~\bibnamefont
  {Ruostekoski}}, \bibinfo {author} {\bibfnamefont {J.}~\bibnamefont
  {Javanainen}}, \bibinfo {author} {\bibfnamefont {R.}~\bibnamefont
  {Bourgain}}, \bibinfo {author} {\bibfnamefont {S.}~\bibnamefont {Jennewein}},
  \bibinfo {author} {\bibfnamefont {Y.~R.~P.} \bibnamefont {Sortais}},  and
  \bibinfo {author} {\bibfnamefont {A.}~\bibnamefont {Browaeys}}, }\bibfield
  {title} {\enquote {\bibinfo {title} {Optical resonance shifts in the
  fluorescence of thermal and cold atomic gases},} }\href {\doibase
  10.1103/PhysRevLett.116.183601} {\bibfield  {journal} {\bibinfo  {journal}
  {Phys. Rev. Lett.} }\textbf {\bibinfo {volume} {116}}, \bibinfo {pages}
  {183601} (\bibinfo {year} {2016})}\BibitemShut {NoStop}%
\bibitem [{\citenamefont {Lewenstein} \emph {et~al.}(1994)\citenamefont
  {Lewenstein}, \citenamefont {You}, \citenamefont {Cooper}, and \citenamefont
  {Burnett}}]{Lewenstein94}%
  \BibitemOpen
  \bibfield  {author} {\bibinfo {author} {\bibfnamefont {M.}~\bibnamefont
  {Lewenstein}}, \bibinfo {author} {\bibfnamefont {L.}~\bibnamefont {You}},
  \bibinfo {author} {\bibfnamefont {J.}~\bibnamefont {Cooper}},  and \bibinfo
  {author} {\bibfnamefont {K.}~\bibnamefont {Burnett}}, }\bibfield  {title}
  {\enquote {\bibinfo {title} {Quantum field theory of atoms interacting with
  photons: Foundations},} }\href {\doibase 10.1103/PhysRevA.50.2207} {\bibfield
   {journal} {\bibinfo  {journal} {Phys. Rev. A} }\textbf {\bibinfo {volume}
  {50}}, \bibinfo {pages} {2207--2231} (\bibinfo {year} {1994})}\BibitemShut
  {NoStop}%
\bibitem [{\citenamefont {Ruostekoski} and \citenamefont
  {Javanainen}(1997)}]{Ruostekoski97}%
  \BibitemOpen
  \bibfield  {author} {\bibinfo {author} {\bibfnamefont {J.}~\bibnamefont
  {Ruostekoski}} and \bibinfo {author} {\bibfnamefont {J.}~\bibnamefont
  {Javanainen}}, }\bibfield  {title} {\enquote {\bibinfo {title} {Quantum field
  theory of cooperative atom response: Low light intensity},} }\href {\doibase
  10.1103/PhysRevA.55.513} {\bibfield  {journal} {\bibinfo  {journal} {Phys.
  Rev. A} }\textbf {\bibinfo {volume} {55}}, \bibinfo {pages} {513--526}
  (\bibinfo {year} {1997})}\BibitemShut {NoStop}%
\bibitem [{\citenamefont {Javanainen} \emph {et~al.}(1999)\citenamefont
  {Javanainen}, \citenamefont {Ruostekoski}, \citenamefont {Vestergaard}, and
  \citenamefont {Francis}}]{Javanainen99}%
  \BibitemOpen
  \bibfield  {author} {\bibinfo {author} {\bibfnamefont {J.}~\bibnamefont
  {Javanainen}}, \bibinfo {author} {\bibfnamefont {J.}~\bibnamefont
  {Ruostekoski}}, \bibinfo {author} {\bibfnamefont {B.}~\bibnamefont
  {Vestergaard}},  and \bibinfo {author} {\bibfnamefont {M.~R.} \bibnamefont
  {Francis}}, }\bibfield  {title} {\enquote {\bibinfo {title} {One-dimensional
  modeling of light propagation in dense and degenerate samples},} }\href
  {\doibase 10.1103/PhysRevA.59.649} {\bibfield  {journal} {\bibinfo  {journal}
  {Phys. Rev. A} }\textbf {\bibinfo {volume} {59}}, \bibinfo {pages} {649--666}
  (\bibinfo {year} {1999})}\BibitemShut {NoStop}%
\bibitem [{\citenamefont {Lee} \emph {et~al.}(2016)\citenamefont {Lee},
  \citenamefont {Jenkins}, and \citenamefont {Ruostekoski}}]{Ruostekoski2016}%
  \BibitemOpen
  \bibfield  {author} {\bibinfo {author} {\bibfnamefont {M.~D.} \bibnamefont
  {Lee}}, \bibinfo {author} {\bibfnamefont {S.~D.} \bibnamefont {Jenkins}},
  and \bibinfo {author} {\bibfnamefont {J.}~\bibnamefont {Ruostekoski}},
  }\bibfield  {title} {\enquote {\bibinfo {title} {Stochastic methods for light
  propagation and recurrent scattering in saturated and nonsaturated atomic
  ensembles},} }\href {\doibase 10.1103/PhysRevA.93.063803} {\bibfield
  {journal} {\bibinfo  {journal} {Phys. Rev. A} }\textbf {\bibinfo {volume}
  {93}}, \bibinfo {pages} {063803} (\bibinfo {year} {2016})}\BibitemShut
  {NoStop}%
\bibitem [{\citenamefont {Childs}(1970)}]{Childs1970}%
  \BibitemOpen
  \bibfield  {author} {\bibinfo {author} {\bibfnamefont {W.~J.} \bibnamefont
  {Childs}}, }\bibfield  {title} {\enquote {\bibinfo {title} {Hyperfine
  structure of $^{5}i_{8,7}$ atomic states of ${\mathrm{dy}}^{161,163}$ and the
  ground-state nuclear moments},} }\href {\doibase 10.1103/PhysRevA.2.1692}
  {\bibfield  {journal} {\bibinfo  {journal} {Phys. Rev. A} }\textbf {\bibinfo
  {volume} {2}}, \bibinfo {pages} {1692--1701} (\bibinfo {year}
  {1970})}\BibitemShut {NoStop}%
\bibitem [{\citenamefont {Petersen} \emph {et~al.}(2020)\citenamefont
  {Petersen}, \citenamefont {Tr\"umper}, and \citenamefont
  {Windpassinger}}]{Petersen2020}%
  \BibitemOpen
  \bibfield  {author} {\bibinfo {author} {\bibfnamefont {N.}~\bibnamefont
  {Petersen}}, \bibinfo {author} {\bibfnamefont {M.}~\bibnamefont {Tr\"umper}},
   and \bibinfo {author} {\bibfnamefont {P.}~\bibnamefont {Windpassinger}},
  }\bibfield  {title} {\enquote {\bibinfo {title} {Spectroscopy of the 1001-nm
  transition in atomic dysprosium},} }\href {\doibase
  10.1103/PhysRevA.101.042502} {\bibfield  {journal} {\bibinfo  {journal}
  {Phys. Rev. A} }\textbf {\bibinfo {volume} {101}}, \bibinfo {pages} {042502}
  (\bibinfo {year} {2020})}\BibitemShut {NoStop}%
\bibitem [{\citenamefont {James}(1993)}]{James93}%
  \BibitemOpen
  \bibfield  {author} {\bibinfo {author} {\bibfnamefont {D.~F.~V.} \bibnamefont
  {James}}, }\bibfield  {title} {\enquote {\bibinfo {title} {Frequency shifts
  in spontaneous emission from two interacting atoms},} }\href {\doibase
  10.1103/PhysRevA.47.1336} {\bibfield  {journal} {\bibinfo  {journal} {Phys.
  Rev. A} }\textbf {\bibinfo {volume} {47}}, \bibinfo {pages} {1336--1346}
  (\bibinfo {year} {1993})}\BibitemShut {NoStop}%
\bibitem [{\citenamefont {Lehmberg}(1970)}]{Lehmberg70}%
  \BibitemOpen
  \bibfield  {author} {\bibinfo {author} {\bibfnamefont {R.~H.} \bibnamefont
  {Lehmberg}}, }\bibfield  {title} {\enquote {\bibinfo {title} {Radiation from
  an $n$-atom system. i. general formalism},} }\href {\doibase
  10.1103/PhysRevA.2.883} {\bibfield  {journal} {\bibinfo  {journal} {Phys.
  Rev. A} }\textbf {\bibinfo {volume} {2}}, \bibinfo {pages} {883--888}
  (\bibinfo {year} {1970})}\BibitemShut {NoStop}%
\bibitem [{\citenamefont {M{\"u}hlbauer} \emph {et~al.}(2018)\citenamefont
  {M{\"u}hlbauer}, \citenamefont {Petersen}, \citenamefont {Baumg{\"a}rtner},
  \citenamefont {Maske}, and \citenamefont {Windpassinger}}]{Muhlbauer2018}%
  \BibitemOpen
  \bibfield  {author} {\bibinfo {author} {\bibfnamefont {F.}~\bibnamefont
  {M{\"u}hlbauer}}, \bibinfo {author} {\bibfnamefont {N.}~\bibnamefont
  {Petersen}}, \bibinfo {author} {\bibfnamefont {C.}~\bibnamefont
  {Baumg{\"a}rtner}}, \bibinfo {author} {\bibfnamefont {L.}~\bibnamefont
  {Maske}},  and \bibinfo {author} {\bibfnamefont {P.}~\bibnamefont
  {Windpassinger}}, }\bibfield  {title} {\enquote {\bibinfo {title} {Systematic
  optimization of laser cooling of dysprosium},} }\href {\doibase
  10.1007/s00340-018-6981-2} {\bibfield  {journal} {\bibinfo  {journal}
  {Applied Physics B} }\textbf {\bibinfo {volume} {124}}, \bibinfo {pages}
  {120} (\bibinfo {year} {2018})}\BibitemShut {NoStop}%
\bibitem [{\citenamefont {Lu} \emph {et~al.}(2011)\citenamefont {Lu},
  \citenamefont {Burdick}, \citenamefont {Youn}, and \citenamefont
  {Lev}}]{benlevbec2011}%
  \BibitemOpen
  \bibfield  {author} {\bibinfo {author} {\bibfnamefont {M.}~\bibnamefont
  {Lu}}, \bibinfo {author} {\bibfnamefont {N.~Q.} \bibnamefont {Burdick}},
  \bibinfo {author} {\bibfnamefont {S.~H.} \bibnamefont {Youn}},  and \bibinfo
  {author} {\bibfnamefont {B.~L.} \bibnamefont {Lev}}, }\bibfield  {title}
  {\enquote {\bibinfo {title} {Strongly dipolar bose-einstein condensate of
  dysprosium},} }\href {\doibase 10.1103/PhysRevLett.107.190401} {\bibfield
  {journal} {\bibinfo  {journal} {Phys. Rev. Lett.} }\textbf {\bibinfo {volume}
  {107}}, \bibinfo {pages} {190401} (\bibinfo {year} {2011})}\BibitemShut
  {NoStop}%
\bibitem [{\citenamefont {Lu} \emph {et~al.}(2012)\citenamefont {Lu},
  \citenamefont {Burdick}, and \citenamefont {Lev}}]{benlevfermi2012}%
  \BibitemOpen
  \bibfield  {author} {\bibinfo {author} {\bibfnamefont {M.}~\bibnamefont
  {Lu}}, \bibinfo {author} {\bibfnamefont {N.~Q.} \bibnamefont {Burdick}},  and
  \bibinfo {author} {\bibfnamefont {B.~L.} \bibnamefont {Lev}}, }\bibfield
  {title} {\enquote {\bibinfo {title} {Quantum degenerate dipolar fermi gas},}
  }\href {\doibase 10.1103/PhysRevLett.108.215301} {\bibfield  {journal}
  {\bibinfo  {journal} {Phys. Rev. Lett.} }\textbf {\bibinfo {volume} {108}},
  \bibinfo {pages} {215301} (\bibinfo {year} {2012})}\BibitemShut {NoStop}%
\bibitem [{\citenamefont {Frisch} \emph {et~al.}(2012)\citenamefont {Frisch},
  \citenamefont {Aikawa}, \citenamefont {Mark}, \citenamefont {Rietzler},
  \citenamefont {Schindler}, \citenamefont {Zupani\ifmmode~\check{c}\else
  \v{c}\fi{}}, \citenamefont {Grimm}, and \citenamefont
  {Ferlaino}}]{ferlaino2012}%
  \BibitemOpen
  \bibfield  {author} {\bibinfo {author} {\bibfnamefont {A.}~\bibnamefont
  {Frisch}}, \bibinfo {author} {\bibfnamefont {K.}~\bibnamefont {Aikawa}},
  \bibinfo {author} {\bibfnamefont {M.}~\bibnamefont {Mark}}, \bibinfo {author}
  {\bibfnamefont {A.}~\bibnamefont {Rietzler}}, \bibinfo {author}
  {\bibfnamefont {J.}~\bibnamefont {Schindler}}, \bibinfo {author}
  {\bibfnamefont {E.}~\bibnamefont {Zupani\ifmmode~\check{c}\else \v{c}\fi{}}},
  \bibinfo {author} {\bibfnamefont {R.}~\bibnamefont {Grimm}},  and \bibinfo
  {author} {\bibfnamefont {F.}~\bibnamefont {Ferlaino}}, }\bibfield  {title}
  {\enquote {\bibinfo {title} {Narrow-line magneto-optical trap for erbium},}
  }\href {\doibase 10.1103/PhysRevA.85.051401} {\bibfield  {journal} {\bibinfo
  {journal} {Phys. Rev. A} }\textbf {\bibinfo {volume} {85}}, \bibinfo {pages}
  {051401} (\bibinfo {year} {2012})}\BibitemShut {NoStop}%
\bibitem [{\citenamefont {Maier} \emph {et~al.}(2014)\citenamefont {Maier},
  \citenamefont {Kadau}, \citenamefont {Schmitt}, \citenamefont {Griesmaier},
  and \citenamefont {Pfau}}]{pfau2014}%
  \BibitemOpen
  \bibfield  {author} {\bibinfo {author} {\bibfnamefont {T.}~\bibnamefont
  {Maier}}, \bibinfo {author} {\bibfnamefont {H.}~\bibnamefont {Kadau}},
  \bibinfo {author} {\bibfnamefont {M.}~\bibnamefont {Schmitt}}, \bibinfo
  {author} {\bibfnamefont {A.}~\bibnamefont {Griesmaier}},  and \bibinfo
  {author} {\bibfnamefont {T.}~\bibnamefont {Pfau}}, }\bibfield  {title}
  {\enquote {\bibinfo {title} {Narrow-line magneto-optical trap for dysprosium
  atoms},} }\href {\doibase 10.1364/OL.39.003138} {\bibfield  {journal}
  {\bibinfo  {journal} {Opt. Lett.} }\textbf {\bibinfo {volume} {39}}, \bibinfo
  {pages} {3138--3141} (\bibinfo {year} {2014})}\BibitemShut {NoStop}%
\bibitem [{\citenamefont {Gustavson} \emph {et~al.}(2001)\citenamefont
  {Gustavson}, \citenamefont {Chikkatur}, \citenamefont {Leanhardt},
  \citenamefont {G\"orlitz}, \citenamefont {Gupta}, \citenamefont {Pritchard},
  and \citenamefont {Ketterle}}]{ketterletransport2001}%
  \BibitemOpen
  \bibfield  {author} {\bibinfo {author} {\bibfnamefont {T.~L.} \bibnamefont
  {Gustavson}}, \bibinfo {author} {\bibfnamefont {A.~P.} \bibnamefont
  {Chikkatur}}, \bibinfo {author} {\bibfnamefont {A.~E.} \bibnamefont
  {Leanhardt}}, \bibinfo {author} {\bibfnamefont {A.}~\bibnamefont
  {G\"orlitz}}, \bibinfo {author} {\bibfnamefont {S.}~\bibnamefont {Gupta}},
  \bibinfo {author} {\bibfnamefont {D.~E.} \bibnamefont {Pritchard}},  and
  \bibinfo {author} {\bibfnamefont {W.}~\bibnamefont {Ketterle}}, }\bibfield
  {title} {\enquote {\bibinfo {title} {Transport of bose-einstein condensates
  with optical tweezers},} }\href {\doibase 10.1103/PhysRevLett.88.020401}
  {\bibfield  {journal} {\bibinfo  {journal} {Phys. Rev. Lett.} }\textbf
  {\bibinfo {volume} {88}}, \bibinfo {pages} {020401} (\bibinfo {year}
  {2001})}\BibitemShut {NoStop}%
\bibitem [{\citenamefont {Alt}(2002)}]{Alt2001}%
  \BibitemOpen
  \bibfield  {author} {\bibinfo {author} {\bibfnamefont {W.}~\bibnamefont
  {Alt}}, }\bibfield  {title} {\enquote {\bibinfo {title} {An objective lens
  for efficient fluorescence detection of single atoms},} }\href {\doibase
  https://doi.org/10.1078/0030-4026-00133} {\bibfield  {journal} {\bibinfo
  {journal} {Optik} }\textbf {\bibinfo {volume} {113}}, \bibinfo {pages}
  {142--144} (\bibinfo {year} {2002})}\BibitemShut {NoStop}%
\bibitem [{\citenamefont {Li} \emph {et~al.}(2018)\citenamefont {Li},
  \citenamefont {Zhou}, \citenamefont {Ke}, \citenamefont {Xu}, \citenamefont
  {He}, \citenamefont {Wang}, and \citenamefont {Zhan}}]{Liobj2018}%
  \BibitemOpen
  \bibfield  {author} {\bibinfo {author} {\bibfnamefont {X.}~\bibnamefont
  {Li}}, \bibinfo {author} {\bibfnamefont {F.}~\bibnamefont {Zhou}}, \bibinfo
  {author} {\bibfnamefont {M.}~\bibnamefont {Ke}}, \bibinfo {author}
  {\bibfnamefont {P.}~\bibnamefont {Xu}}, \bibinfo {author} {\bibfnamefont
  {X.-D.} \bibnamefont {He}}, \bibinfo {author} {\bibfnamefont
  {J.}~\bibnamefont {Wang}},  and \bibinfo {author} {\bibfnamefont {M.-S.}
  \bibnamefont {Zhan}}, }\bibfield  {title} {\enquote {\bibinfo {title}
  {High-resolution ex vacuo objective for cold atom experiments},} }\href
  {\doibase 10.1364/AO.57.007584} {\bibfield  {journal} {\bibinfo  {journal}
  {Appl. Opt.} }\textbf {\bibinfo {volume} {57}}, \bibinfo {pages} {7584--7590}
  (\bibinfo {year} {2018})}\BibitemShut {NoStop}%
\end{thebibliography}%

\onecolumngrid

\appendix
\newpage
\section{Magnetic interaction terms in light matter Hamiltonian from first principles}\label{sec:derivation}

We will be using the following definitions for the gauge, electric and magnetic field operators
\begin{align}\label{eq:app_fields}
  \hat{\bm A}&=\sum_{\bm k,\bm \epsilon}g_k\epsilon_{\bm k}\left(\hat a_{\bm k,\bm \epsilon}\me^{\iu\bm k\bm R}+\hat a_{\bm k,\bm \epsilon}^\dagger\me^{-\iu\bm k\bm R}\right)\\
  \hat{\bm E}&=\iu\sum_{\bm k,\bm \epsilon}g_k\omega_k\epsilon_{\bm k}\left(\hat a_{\bm k,\bm \epsilon}\me^{\iu\bm k\bm R}-\hat a_{\bm k,\bm \epsilon}^\dagger\me^{-\iu\bm k\bm R}\right)\\
  \hat{\bm B}&=\iu\sum_{\bm k,\bm \epsilon}\left(\bm k\cross\epsilon_{\bm k}\right)g_k\left(\hat a_{\bm k,\bm \epsilon}\me^{\iu\bm k\bm R}-\hat a_{\bm k,\bm \epsilon}^\dagger\me^{-\iu\bm k\bm R}\right).
\end{align}
with $g_k=1/\sqrt{2\omega_k \mathcal{V}\epsilon_0}$ photon coupling strength with photonic operators as defined in the main text.

If we omit terms where the square of the gauge field appears and the term where the atom core momentum is coupled to the gauge field (amounting to a Born-Oppenheimer and subsequently a rotating wave approximation), the simplified Hamiltonian is composed of three parts, written in the Coulomb gauge

\begin{align}
  \mathcal{H}&\approx\mathcal{H}_0+\mathcal{H}_{\text{af}}+\mathcal{H}_{\text{dipole}}\\
  \mathcal{H}_0&=\sum_{i}^N\frac{\hat{\bm P}_i^2}{2M}+\frac{\hat{\bm p}_i^2}{2m}+V_{ec}(\bm R_i,\bm r_i)\\
  \mathcal{H}_{af}&=\frac{1}{m}\sum_{i}^N\hat{\bm{A}}(\bm r_i)\cdot \hat{\bm p}_i\\
  \mathcal{H}_{\text{dipole}}&=\frac{1}{2}\sum_{i\neq j}\sum\frac{\bm d_i\cdot\bm d_j-3(\bm R_{ij}\bm d_i)(\bm R_{ij}\cdot\bm d_j)}{R^3_{ij}}.
\end{align}

where we have already performed a dipole approximation for the Coulomb interaction between different atoms leading to the explicit form of $\mathcal{H}_{\text{dipole}}$. The Hamiltonian $\mathcal{H}_{af}$ is the canonical coupling between the electronic degree of freedom and the transverse modes of the light-field. We have not completed the dipolar approximation in the gauge field which still contains the position of the electron $\bm r_i$ which is a quantum operator in this description. The Hamiltonian $\mathcal H_0$ contains the electron kinetic energy, the center of mass kinetic energy and the interaction between the electron and the core. The electron kinetic energy and the core-electron potential give rise to the level structure of the atom. We consider a Born-Oppenheimer picture where $\bm R_i$ represents the center of mass momentum.

Instead of making the usual dipolar approximation for the gauge field, we consider one order higher in the Taylor approximation of the electron position around the center of mass position $\me^{\iu\bm k\bm r_i}\approx\me^{\iu\bm k\bm R_i}\left[1+\iu\bm k\bm x_i\right]$ with the distance vector $\bm x_i=\bm r_i-\bm R_i$. The atom field Hamiltonian then splits into two contributions

\begin{equation}
  \begin{split}
    \mathcal{H}_{af}&=\frac{1}{m}\sum_{i}^N\hat{\bm p}_i\cdot\hat{\bm{A}}(\bm R_i)
    +\iu\sum_{\bm k,\bm \epsilon}g_k\left(\hat{\bm p}_i\cdot\epsilon_{\bm k}\right)(\bm k\bm x_i)\left(\hat a_{\bm k,\bm \epsilon}\me^{\iu\bm k\bm R_i}-\hat a_{\bm k,\bm \epsilon}^\dagger\me^{-\iu\bm k\bm R_i}\right).
  \end{split}
\end{equation}

The first term is the simple dipole approximation which leads to the coupled dipole model after some additional approximations and a gauge transformation. The second term is of higher order and contains the magnetic and quadrupole electric moment as we will see. In order to see that the double dot product indeed yields the magnetic and quadrupole term, it needs to be rearranged
\begin{equation}
  \begin{split}
    (\hat{\bm p}_i\cdot\bm \epsilon_{\bm k})(\bm k\cdot\bm x_i)&=(\hat{\bm p}_i\cdot \bm k)(\bm x_i\cdot \bm \epsilon_{\bm k})+(\bm x_i\cross\hat{\bm p}_i)(\bm k\cross\epsilon_{\bm k})\\
    &=\frac{1}{2}\left((\hat{\bm p}_i\cdot\bm \epsilon_{\bm k})(\bm k\cdot\bm x_i)+(\hat{\bm p}_i\cdot \bm k)(\bm x_i\cdot \bm \epsilon_{\bm k})\right)
    +\frac{1}{2}\hat{\bm L}(\bm k\cross\epsilon_{\bm k})
  \end{split}
\end{equation}

The first term is the electronic quadrupole term whereas the second term contains the angular momentum implying that it is indeed the magnetic coupling. Summarizing the calculations above, the terms that must be added to the Hamiltonian if the dipole approximation is performed to higher order are

\begin{equation}
  \begin{split}
    \mathcal{H}_m&=\sum_i^N\frac{\bm{\hat L}_i}{2m}\iu\sum_{\bm k,\bm \epsilon}g_k(\bm k\cross\bm \epsilon_{\bm k})\left(\hat a_{\bm k,\bm \epsilon}\me^{\iu\bm k\bm R_i}-\hat a_{\bm k,\bm \epsilon}^\dagger\me^{-\iu\bm k\bm R_i}\right)
    =\sum_i^N\hat{\bm\mu}_i\hat{\bm B}(\bm R_i)\\
    \mathcal{H}_q&=\sum_{i,\bm k,\bm \epsilon}\frac{\iu \abs{\bm k}}{2m}(p_{\bm k}r_{\bm \epsilon}+p_{\bm \epsilon}r_{\bm k})g_k\left(\hat a_{\bm k,\bm \epsilon}\me^{\iu\bm k\bm R_i}-\hat a_{\bm k,\bm \epsilon}^\dagger\me^{-\iu\bm k\bm R_i}\right)
  \end{split}
\end{equation}

Here we avoid complications with spin degrees of freedom by making the association $\hat{\bm\mu}_i=-\frac{\bm{\hat L}_i}{2m}$ and noting that spin degrees of freedom need to be explicitly inserted into the Hamiltonian which ultimately change the definition of $\hat{\bm\mu}_i$. Since we go into a two level picture this is however already taken care of and we need not consider these complications in detail.

\section{Representation of the Hamiltonian in the two-level approximation}
\label{sec:appendixb}

We perform a two-level approximation in which we assume there exist two relevant levels per atom, the ground state $\ket{g_i}$ and the excited state $\ket{e_i}$ separated by a frequency $\omega_0$. Taking the projection of the momentum  and magnetic moment operator onto this subspace, we can explicilty write the Hamiltonian using the matrix elements within this subspace. One relevant point is here that the momentum operator has no diagonal components whereas the magnetic moment operator has no such symmetry restriction.

Omitting the quadrupole term, the atom-field Hamiltonian then becomes
\begin{equation}
  \begin{split}
    \mathcal{H}_{\text{af}}&=\iu\sum_i^N\left(\bm d_i\sigma_i-\bm d^*_i\sigma_i^\dagger\right)\hat{\bm A}(\bm R_i)
    +\left(\bm \mu_i\sigma_i+\bm \mu^*_i\sigma_i^\dagger+\bm\mu_{i,e}\sigma^\dagger\sigma+\bm\mu_{i,g}\sigma\sigma^\dagger\right)\hat{\bm B}(\bm R_i).
  \end{split}
\end{equation}
with $\bm d_i=\matrixelement{g_i}{\hat x_i}{e_i}$ and $\bm\mu_{i,e}=\matrixelement{g_i}{\hat{\bm\mu}_i}{e_i}$ for the transition dipoles and $\bm\mu_{i,e}=\matrixelement{e_i}{\hat{\bm\mu}_i}{e_i}$, $\bm\mu_{i,g}=\matrixelement{g_i}{\hat{\bm\mu}_i}{g_i}$ for the static magnetic dipole moment. The sigma matrices are transition operators of the type $\sigma_i=\ket{g_i}\bra{e_i}$.

\section{Unitary transformation into dipolar gauge}
\label{sec:appendixc}

The next step in deriving the light-matter Hamiltonian is the gauge transformation

\begin{equation}
  U=\exp\left(-\iu\sum_i^N\int_{\bm r_i}^{\bm R_i}\diff{\bm r}\hat{\bm A}(\bm r)\right)
\end{equation}

We note here that the gauge field operator commutes with the magnetic field operator $\commutator{\hat{\bm A}(\bm r)}{\hat{\bm B}(\bm r')}=0$ so that $H_m$ is not affected by this transformation. The effect on the dipole electronic transition and the static dipole $\mathcal{H}_{\text{dipole}}$ is as usual so that, after also performing a rotating wave approximation, the Hamiltonian is
\begin{equation}\label{eq:light_matter_mag}
  \begin{split}
    \mathcal{H}&=\omega\sum_i\sigma_i^\dagger\sigma_i
    +\sum_{\bm k,\bm \epsilon}\omega_k\hat a_{\bm k,\bm \epsilon}^\dagger\hat a_{\bm k,\bm \epsilon}
    +\sum_i\hat{\bm d}_i\cdot\hat{\bm E}(\bm R_i)+\left(\hat{\bm \mu}_{t,i}+\hat{\bm \mu}_{s,i}\right)\cdot\hat{\bm B}(\bm R_i).
  \end{split}
\end{equation}
with definitions $\hat{\bm d}_j=\bm d\sigma_j+\bm d^*\sigma_j^\dagger$, $\hat{\bm \mu}_{t,i}=\bm \mu_i\sigma_i+\bm \mu^*_i\sigma_i^\dagger$ and $\hat{\bm \mu}_{s,i}=\bm\mu_{i,e}\sigma^\dagger\sigma+\bm\mu_{i,g}\sigma\sigma^\dagger$.

\section{Elimination of photonic degrees of freedom from Hamiltonian}
\label{sec:elimination}

The last step in determining the effective system dynamics is to perform an adiabatic elimination of the light modes for this Hamiltonian Eq.~\eqref{eq:light_matter_mag}. The first step is to determine the equation of motion for the photonic operators
\begin{equation}
  \begin{split}
    \dv{\hat{a}_{\bm{k},\bm{\epsilon}}}{t}&=-\iu\omega_{k}\hat{a}_{\bm{k},\bm{\epsilon}}+\sum_ig_{k}\omega_k(\hat{\bm{d}}_i\cdot\bm{\epsilon}_{\bm{k}})\me^{-\iu\bm{k}\hat{\bm{R}}_i}
    +\sum_ig_{k}\left(\left[\hat{\bm\mu}_{t,i}+\hat{\bm\mu}_{s,i}\right]\cdot\left[\bm k\cross\bm{\epsilon}_{\bm{k}}\right]\right)\me^{-\iu\bm{k}\hat{\bm{R}}_i}
  \end{split}
\end{equation}
which can then be integrated to yield
\begin{equation}
  \begin{split}
    \hat{a}_{\bm{k},\bm{\epsilon}}(t)&=\hat{a}_{\bm{k},\bm{\epsilon}}(0)\me^{-\iu\omega_kt}
    +\int_0^t\diff{s}\me^{-\iu\omega_k(t-s)}
    \sum_ig_{k}\omega_k(\hat{\bm{d}}_i(s)\cdot\bm{\epsilon}_{\bm{k}})\me^{-\iu\bm{k}\hat{\bm{R}}_i}\\
    &+\int_0^t\diff{s}\me^{-\iu\omega_k(t-s)}
    \sum_ig_{k}\left(\left[\hat{\bm\mu}_{t,i}(s)+\hat{\bm\mu}_{s,i}(s)\right]\cdot\left[\bm k\cross\bm{\epsilon}_{\bm{k}}\right]\right)\me^{-\iu\bm{k}\hat{\bm{R}}_i}.
  \end{split}
\end{equation}

A simplification for the calculation that follows is to consider negative and positive frequency components of the magnetic and electric field in particular
\begin{equation}
  \label{eq:posnegfield}
  \begin{split}
    \hat{\bm B}^\pm&=\pm\iu\sum_{\bm k,\bm \epsilon}\left(\bm k\cross\epsilon_{\bm k}\right)g_k\hat a_{\bm k,\bm \epsilon}\me^{\pm\iu\bm k\bm R}\\
    \hat{\bm E}^\pm&=\pm\iu\sum_{\bm k,\bm \epsilon}g_k\omega_k\epsilon_{\bm k}\hat a_{\bm k,\bm \epsilon}\me^{\pm\iu\bm k\bm R}
  \end{split}
\end{equation}
so that we only have to consider the equation of motion for one of the photonic operators. To shorten the notation we also introduce the total magnetic dipole operator $\hat{\bm \mu}_i(t)=\hat{\bm\mu}_{t,i}(t)+\hat{\bm\mu}_{s,i}(t)$ leading to expressions for the positive field components of the electric and magnetic field
\begin{equation}
  \begin{split}
    \hat{\bm E}^+(\bm R_i)&=\hat{\bm E}_0^+(\bm R_i)
    +\iu\sum_{\bm k,\bm \epsilon,j}(g_k\omega_k)^2\me^{\iu\bm k\bm R_{ij}}\epsilon_{\bm k}
    \int_0^t\diff{s}\me^{-\iu\omega_k(t-s)}(\hat{\bm{d}}_j(s)\cdot\bm{\epsilon}_{\bm{k}})\\
    &+\iu\sum_{\bm k,\bm \epsilon,j}g^2_k\omega_k\me^{\iu\bm k\bm R_{ij}}\epsilon_{\bm k}
    \int_0^t\diff{s}\me^{-\iu\omega_k(t-s)}\left(\hat{\bm \mu}_i(s)\cdot\left[\bm k\cross\bm{\epsilon}_{\bm{k}}\right]\right)\\
    \hat{\bm B}^+(\bm R_i)&=\hat{\bm B}_0^+(\bm R_i)
    +\iu\sum_{\bm k,\bm \epsilon, j}\left(\bm k\cross\epsilon_{\bm k}\right)g^2_k\omega_k\me^{\iu\bm k\bm R_{ij}}
    \int_0^t\diff{s}\me^{-\iu\omega_k(t-s)}(\hat{\bm{d}}_j(s)\cdot\bm{\epsilon}_{\bm{k}})\\
    &+\iu\sum_{\bm k,\bm \epsilon, j}\left(\bm k\cross\epsilon_{\bm k}\right)g^2_k\me^{\iu\bm k\bm R_{ij}}
    \int_0^t\diff{s}\me^{-\iu\omega_k(t-s)}\left(\hat{\bm \mu}_i(s)\cdot\left[\bm k\cross\bm{\epsilon}_{\bm{k}}\right]\right)
  \end{split}
\end{equation}
where the zero subscript indicates that the photonic operators in the field expression were taken at time zero. Thus, both fields have similar terms. First, the vacuum time evolution, then the interaction with the proper dipole moment and then a cross term between the magnetic and electric components of the interaction.

Using the usual results for sums over polarization vectors since they form a orthonormal basis with the normalized wavevector, the cross terms can be combined into
\begin{equation}
  \begin{split}
    \sum_{\bm\epsilon}\left((\hat{\bm d}_i(t)\cdot \bm\epsilon_{\bm k})\hat{\bm \mu}_i(s)\cdot\left[\bm k\cross\bm{\epsilon}_{\bm{k}}\right]
      +\hat{\bm \mu}_i(t)\cdot\left[\bm k\cross\bm{\epsilon}_{\bm{k}}\right]
      (\hat{\bm{d}}_j(s)\cdot\bm{\epsilon}_{\bm{k}})
    \right)
    =\left(\hat{\bm \mu}_i(t)\cross\hat{\bm d}_i(s)-\hat{\bm d}_i(t)\cross\hat{\bm \mu}_i(s)\right)\cdot\bm k
  \end{split}
\end{equation}
which is linear in $\bm{k}$, such that the negative components of the sum of $\hat{\bm E}^+(\bm R_i)$ cancel the positive wave vectors in $\hat{\bm E}^-(\bm R_i)$ since the dipole operator is hermitian and the polarization has been chosen real. The same goes for the magnetic field, canceling the cross terms.

After the cross-terms are cancelled, the total magnetic and electric field operators can be obtained from the positive frequency components by adding the hermitian conjugate so that after performing the usual summation over the polarization vectors
\begin{align}
  \hat{\bm E}(\bm R_i)&=\hat{\bm E}_0(\bm R_i)
                        +\iu\sum_{\bm k,j}(g_k\omega_k)^2\me^{\iu\bm k\bm R_{ij}}\int_0^t\diff{s}\me^{-\iu\omega_k(t-s)}
                        \cdot\left[\hat{\bm{d}}_j(s)-\frac{\bm{k}(\hat{\bm{d}}_j(s)\cdot\bm{k})}{k^2}\right]+\hc\\
  \hat{\bm B}(\bm R_i)&=\hat{\bm B}_0(\bm R_i)
                        +\iu\sum_{\bm k,j}(g_k\omega_k)^2\me^{\iu\bm k\bm R_{ij}}\int_0^t\diff{s}\me^{-\iu\omega_k(t-s)}
                        \cdot\left[\hat{\bm \mu}_i(s)-\frac{\bm{k}(\hat{\bm \mu}_i(s)\cdot\bm{k})}{k^2}\right]+\hc
\end{align}
so that the magnetic and electric fields have the same structure. Now, the continuum limit is taken for the electromagnetic field modes $\frac{1}{V}\sum_{\bm k}\rightarrow\frac{1}{(2\pi)^3}\int\diff{\bm k}$ and the angular integral is performed in spherical coordinates so that, defining the dipolar Green's function
\begin{equation}
  \label{eq:greens_function}
  \begin{split}
  \mathbf{G}_k(\mathbf{R})&=\left(\mathds{1}+\frac{1}{k^2} \nabla\otimes\nabla\right)\frac{\me^{\mi kR}}{4\pi R}\\
  &=\frac{\me^{\mi k R}}{4\pi k^2}\left[\left(\frac{k^2}{R}+\frac{\mi k}{R^2}-\frac{1}{R^3}\right)\mathds{1}+\left(-\frac{k^2}{R}-\frac{3\mi k }{R^2}+\frac{3}{R^3}\right)\frac{\mathbf{R}\otimes\mathbf{R}}{R^2}\right]
\end{split}
\end{equation}
with the usual shorthand $\bm G(\bm R)=\bm G_{k_0}(\bm R)$ and the decomposition $\bm G_k(\bm R)=\bm \Omega_k(\bm R)-\iu\bm \Gamma_k(\bm R)$. The expressions for the electric and magnetic field become
\begin{align}
  \hat{\bm E}(\bm R_i)&=\hat{\bm E}_0(\bm R_i)
                        +\frac{\iu}{\pi\epsilon_0}\sum_{j}\int_0^\infty\diff{k}\int_0^t\diff{s}k^2\bm\Gamma_{\bm k}(\bm R_{ij})
                        \cdot\hat{\bm{d}}_j(s)\me^{-\iu\omega_k(t-s)}+\hc\\
  \hat{\bm B}(\bm R_i)&=\hat{\bm B}_0(\bm R_i)
                        +\frac{\iu}{\pi\epsilon_0}\sum_{j}\int_0^\infty\diff{k}\int_0^t\diff{s}k^2\bm\Gamma_{\bm k}(\bm R_{ij})
                        \cdot\hat{\bm{\mu}}_j(s)\me^{-\iu\omega_k(t-s)}+\hc.
\end{align}

The key is now that this elimination cannot be performed exactly and one must make a perturbative Ansatz by plugging the time evolution of the dipole operators without light-matter interactions into these integrals. The free-space time evolution of the dipole operators is explicitly
\begin{equation}
  \begin{split}
    \hat{\bm{d}}_j(s)&=\bm d\sigma_j(s)+\bm d^*\sigma_j^\dagger
    =\bm d\sigma_j(t)\me^{-\iu\omega(s-t)}+\bm d^*\sigma_j(t)^\dagger\me^{\iu\omega(s-t)}\\
    \hat{\bm{\mu}}_{t,j}(s)&=\bm \mu_j\sigma_j(s)+\bm \mu^*_j\sigma_j^\dagger(s)
    =\bm \mu_j\sigma_j(t)\me^{-\iu\omega(s-t)}+\bm \mu^*_j\sigma_j^\dagger(t)\me^{\iu\omega(s-t)}\\
    \hat{\bm{\mu}}_{s,j}(s)&=\bm\mu_{j,e}\sigma_j^\dagger(s)\sigma_j(s)+\bm\mu_{j,g}\sigma_j(s)\sigma_j^\dagger(s)
    =\hat{\bm{\mu}}_{s,j}(t).
  \end{split}
\end{equation}

This implies that the fundamental difference between the static and transition dipole moments is given by the frequency they rotate at. All these time evolutions will lead to integrations of the type
\begin{equation}
  \zeta(\omega)=\iu\int_0^\infty\me^{\iu\omega t}\diff{t}=\mathcal{P}\frac{1}{\omega}+\iu\pi\delta(\omega)
\end{equation}
so that, splitting the magnetic field into a transition part $\hat {\bm B}=\hat {\bm B}_t+\hat {\bm B}_s$ while keeping the vacuum contribution in the transition part
\begin{equation}
  \begin{split}
    \hat{\bm E}(\bm R_i)&=\hat{\bm E}_0(\bm R_i)+\frac{1}{\pi\epsilon_0}\sum_{j}\int_0^\infty\diff{k}k^2\bm\Gamma_{k}(\bm R_{ij})
    \cdot\bm d\left(\sigma_j\zeta(\omega_k+\omega)+\sigma_j^\dagger\zeta(\omega_k-\omega)\right)+\hc\\
    \hat{\bm B}_t(\bm R_i)&=\hat{\bm B}_0(\bm R_i)+\frac{1}{\pi\epsilon_0}\sum_{j}\int_0^\infty\diff{k}k^2\bm\Gamma_{k}(\bm R_{ij})
    \cdot\bm \mu_{t,j}\left(\sigma_j\zeta(\omega_k+\omega)+\sigma_j^\dagger\zeta(\omega_k-\omega)\right)+\hc\\
    \hat{\bm B}_s(\bm R_i)&=\frac{1}{\pi\epsilon_0}\sum_{j}\int_0^\infty\diff{k}k^2\bm\Gamma_{k}(\bm R_{ij})
    \cdot\bm \zeta(\omega_k)\left(\bm\mu_{j,e}\sigma_j^\dagger\sigma_j+\bm\mu_{j,g}\sigma_j\sigma_j^\dagger\right)+\hc.
  \end{split}
\end{equation}

The previous statement about the rotational frequencies now becomes clear as the transition fields contain zeta functions of the form $\zeta(\omega_k\pm\omega)$ where the Green's function around $\bm \Gamma_{k_0}(\bm R)$ matters. The static contributions however contain a zeta function of the form $\zeta(\omega_k)$ where $\bm \Gamma_{k=0}(\bm R)$ matters. Since the zeta function essentially only evaluates the Kramers Kronig relations for the complex function $\bm G(\bm R)$, i.e. $(k'^2)\Omega_{k'}(\bm R)=\int\diff{k}\frac{k^2\Gamma_k(\bm R)}{\omega'-\omega}$. Ultimately, this leads to the explicit representation of the fields, using the fact that $\lim_{k\to 0}k^2 G_k(\bm R)=\frac{1}{4\pi R_{ij}^3}\left(3\frac{\bm R_{ij}\bm R_{ij}}{R_{ij}R_{ij}}-1\right)$ so that only the real part remains

\begin{equation}
  \begin{split}
    \hat{\bm E}(\bm R_i)&=\hat{\bm E}_0(\bm R_i)+\frac{k_0^2}{\epsilon_0}\sum_{j}\left[\left(\bm \Gamma(\bm R_{ij})-\iu \bm \Omega(\bm R_{ij})\right)\sigma_j
      +\left(\bm \Gamma(\bm R_{ij})+\iu\bm \Omega(\bm R_{ij})\right)\sigma_j^\dagger\right]\bm d\\
    \hat{\bm B}(\bm R_i)&=\hat{\bm B}_0(\bm R_i)+\frac{k_0^2}{\epsilon_0}\sum_{j}\left[\left(\bm \Gamma(\bm R_{ij})-\iu \bm \Omega(\bm R_{ij})\right)\sigma_j
      +\left(\bm \Gamma(\bm R_{ij})+\iu\bm \Omega(\bm R_{ij})\right)\sigma_j^\dagger\right]\bm \mu_{t}\\
      \hat{\bm B}_s(\bm R_i)&=\frac{\mu_0}{4\pi}\sum_{j}\frac{1}{R_{ij}^3}\left(3\frac{\bm R_{ij}\bm R_{ij}}{R_{ij}R_{ij}}-1\right)\hat{\bm \mu_{s,j}}.
  \end{split}
\end{equation}

The next step is to plug these expressions into the equations of motion, to use the rotating wave approximation and the definitions $\Gamma=k_0^3d^2/(3\pi\hbar\epsilon_0)$ and $\Gamma_{\mu}=k_0^3\mu^2/3\pi\hbar\epsilon_0$ in order to define the Green's function for the matter problem. Indeed, the definitions $f^{(\mu)}_{ij}=\frac{k_0^2}{\epsilon_0}\bm \mu_{i,t}\bm \Omega(\bm R_{ij})\bm \mu_{j,t}$, $f^{(g)}_{ij}=\frac{k_0^2}{\epsilon_0}\bm d_{i}\bm \Omega(\bm R_{ij})\bm d_{j}$, equivalent definitions for the imaginary part and Eq.~\eqref{eq:mag_interaction_form} lead precisely to Eq.~\eqref{eq:effective_hamiltonian}, Eq.~\eqref{eq:mastereq} and Eq.~\eqref{eq:dissipator} after using the rotating wave approximation.

After adding a classical drive just as in the main text, the equations of motion for the transition operators become

\begin{equation}
  \begin{split}
    \dv{\sigma_k}{t}&=-\iu\left[\Delta
      +\delta\hat\omega_k\right]\sigma_k+\iu\sigma_k^z\sum_{i\neq k}\left(g_{ki}^{(d)}+g_{ki}^{(\mu)}\right)\sigma_i
    +\sigma_k^z\sum_i\left(f_{ki}^{(d)}+f_{ki}^{(\mu)}\right)\sigma_i
    -\iu\Omega\me^{\iu\bm k_0\bm R_i}\sigma_k^z+\sigma_{k,\text{in}}
  \end{split}
\end{equation}

with a frequency operator

\begin{equation}
  \label{eq:freqop}
  \delta\hat{\omega}_k=
  \frac{1}{2}\sum_{i\neq k}\left[\left(\Omega_{ik}^{e,e}-\Omega_{ik}^{g,g}\right)
    +\left(\Omega_{ik}^{e,e}+\Omega_{ik}^{g,g}-2\Omega_{ik}^{e,g}\right)\sigma_i^z\right].
\end{equation}

This is a configuration and density-dependent Zeeman shift of the $k$-th atom's frequency.

\section{Weak-excitation limit}

The weak-excitation limit, i.e. $\expval{\sigma^{z}_j\sigma_{j'}}\approx -\expval{\sigma_{j'}}$, the equations of motion for the expectation values of the transition operators become
\begin{equation}
  \begin{split}
    \dv{\expval{\sigma_k}}{t}&=-\iu\left[\Delta
      +\delta\omega_k\right]\expval{\sigma_k}-\iu\sum_{i\neq k}\left(g_{ki}^{(d)}+g_{ki}^{(\mu)}\right)\expval{\sigma_i}
    -\sum_i\left(f_{ki}^{(d)}+f_{ki}^{(\mu)}\right)\expval{\sigma_i}
    +\iu\Omega\me^{\iu\bm k_0\bm R_i}
  \end{split}
\end{equation}
which is written in matrix notation in Eq.~\eqref{eq:linear_pol_eqn}.

\end{document}